\documentclass[fleqn,usenatbib]{mnras}

\usepackage[T1]{fontenc}

\DeclareRobustCommand{\VAN}[3]{#2}
\let\VANthebibliography\thebibliography
\def\thebibliography{\DeclareRobustCommand{\VAN}[3]{##3}\VANthebibliography}

\usepackage{graphicx}	
\usepackage{amsmath}	
\usepackage{amssymb}	
\usepackage{newtxtext,newtxmath}

\title[Extended stellar haloes of the faintest galaxies]{EDGE: The direct link between mass growth history and the extended stellar haloes of the faintest dwarf galaxies}

\author[A. Goater et al.]
{Alex Goater$^{1}$,
Justin I. Read$^{1}$,
Noelia E. D. Noël$^{1}$,
Matthew D. A. Orkney$^{2}$,
Stacy Y. Kim$^{1}$,
Martin P. Rey$^{3}$,
\newauthor
Eric P. Andersson$^{4}$, 
Oscar Agertz$^{5}$, 
Andrew Pontzen$^{6}$,
Roberta Vieliute$^{7}$, 
Dhairya Kataria$^{8}$ and 
Kiah Jeneway$^{9}$
\\
$^1$Department of Physics, University of Surrey, Guildford, GU2 7XH, UK\\
$^2$ICCUB, Universitat de Barcelona, Martí i Franquès 1, 08028 Barcelona, Spain\\
$^3$Sub-department of Astrophysics, University of Oxford, DWB, Keble Road, Oxford, OX1 3RH, UK\\
$^4$American Museum of Natural History, 200 Central Park West, New York, NY 10024, USA\\
$^5$Lund Observatory, Division of Astrophysics, Department of Physics, Lund University, Box 43, SE-221 00 Lund, Sweden\\
$^6$Department of Physics and Astronomy, University College London, London, WC1E 6BT, UK\\
$^7$School of Physics and Astronomy, University of Southampton, Southampton, SO17 1BJ, UK\\
$^8$School of Physics Engineering and Computer Science, University of Hertfordshire, College Lane, Hatfield, AL10 9AB, UK\\
$^9$School of Physics and Astronomy, University of Kent, Canterbury, CT2 7NH, UK}

\date{Accepted XXX. Received YYY; in original form ZZZ}

\pubyear{2022}

\begin{document}
\label{firstpage}
\pagerange{\pageref{firstpage}--\pageref{lastpage}}
\maketitle

\begin{abstract}
Ultra-faint dwarf galaxies (UFDs) are commonly found in close proximity to the Milky Way and other massive spiral galaxies. As such, their projected stellar ellipticity and extended light distributions are often thought to owe to tidal forces. In this paper, we study the projected stellar ellipticities and faint stellar outskirts of tidally isolated ultra-faints drawn from the `Engineering Dwarfs at Galaxy Formation’s Edge’ (EDGE) cosmological simulation suite. Despite their tidal isolation, our simulated dwarfs exhibit a wide range of projected ellipticities ($0.03 < \varepsilon < 0.85$), with many possessing anisotropic extended stellar haloes that mimic tidal tails, but owe instead to late-time accretion of lower mass companions. Furthermore, we find a strong causal relationship between ellipticity and formation time of an UFD, which is robust to a wide variation in the feedback model. We show that the distribution of projected ellipticities in our suite of simulated EDGE dwarfs matches well with that of 21 Local Group dwarf galaxies. Given the ellipticity in EDGE arises from an ex-situ accretion origin, the agreement in shape indicates the ellipticities of some observed dwarfs may also originate from a similar non-tidal scenario. The orbital parameters of these observed dwarfs further support that they are not currently tidally disrupting. If the baryonic content in these galaxies is still tidally intact, then the same may be true for their dark matter content, making these galaxies in our Local Group pristine laboratories for testing dark matter and galaxy formation models.

\end{abstract}

\begin{keywords}
galaxies: dwarf; galaxies: formation; galaxies: stellar content; galaxies: structure
\end{keywords}



\section{Introduction}\label{Introduction}

The Local Group (hereafter LG) of galaxies offers an excellent laboratory to constrain the $\Lambda$CDM paradigm. The satellite systems orbiting the Milky Way (hereafter MW) allow us to investigate the processes and feedback effects governing galaxy formation and evolution in exquisite detail. In particular, the LG is host to the smallest galaxies known to date, the ultra-faint dwarfs (hereafter UFDs). With the faintest containing a few thousand stars and the brightest having a luminosity of $\sim 10^5$\,L$_\odot$, UFDs represent the extreme lower limit of the galaxy luminosity function \citep{simon2019}, carrying key evidence that can shed light on fundamental galactic processes \citep{battaglia2013, simon2019, agertz2020, sales2022}. These systems are the oldest, most chemically primitive \citep{bromm2011, kirby2013, frebel2014, chiti2021} and most dark matter dominated \citep{brown2014, bb2017, zoutendijk2021, battaglia2022, collins2022} systems in the Universe and, hence, make excellent laboratories to constrain the nature of the mysterious dark matter.

The first major milestone in the search for UFDs was brought about by the advent of digital surveys, such as the Sloan Digital Sky Survey (SDSS), where searches for these faint systems in our LG were completed up to a surface brightness limit of 25.5 mag arcsec$^{-2}$ \citep{whiting2007, koposov2008, willman2005a, belokurov2007, belokurov2008, belokurov2009, belokurov2010, mcconachie2012, belokurov2013}. Thanks to deep imaging and spectroscopy with modern telescopes such as the Dark Energy Camera (DECam) \citep{decam2015}, the past decade has seen an explosion in the number of faint dwarf galaxies discovered, with 68 now known around the MW, 9 around M31 and 15 around galaxies beyond the LG. These newly discovered UFDs in our LG, with solar luminosities $\lesssim 10^5 L_\odot$, have vastly improved our understanding of these systems \citep{betchol2015, koposov2015, koposov2018, dw2015, dw2016, martin2015, kim2015b, kim2015a, kj2015, Laevens2015, torrealba2016a, torrealba2018, homma2016, Homma2018, simon2019, dw2021, cerny2021, collinspegV, sand2022}.
\newline\indent
While UFDs offer the promise of unique constraints on galaxy formation models \citep{collins2022} and the nature of dark matter \citep{simon2019}, this is made more challenging by the potential impact of tidal forces from nearby host galaxies like the MW and M31 \citep{read2006c, lokas2013, collins2017, mazzarini2020}. These larger systems can severely disrupt the environment of nearby smaller galaxies by stripping their dark matter content and then their stellar content, thus deforming the structure of the latter. Evidence for tides has been claimed for many nearby UFDs based on their extended light profiles \citep{li2018, mutlu-pakdil2019, pozo2022}, apparent tidal features \citep{munoz2010, munoz2012}, velocity gradients in their outermost stars \citep{sand2017} and/or constraints on their orbits \citep{kupper2017}. However, the latest constraints on UFD orbits suggest that a number are tidally isolated at present \citep{simon2018, fritz2018, mcconachie2020, pace2022}. A notable example is Tucana II, a LG UFD located at a distance of $\sim58$ kpc away from the MW \citep{pace2022}, which exhibits multiple features characteristic of tidal isolation \citep{chiti2021}. \citet{chiti2021, chiti2022} have recently reported member stars anisotropically distributed around Tucana II, up to nine half-light radii from its galactic centre. These member stars reach out to and even extend past our calculated tidal radius estimates of Tucana II, $r_{\textnormal{t}} \approx 0.76$ kpc. This indicates that there are plenty of stars not of tidal origin in the region $r_{\textnormal{1/2}} < R < r_{\textnormal{t}}$, where $r_\textnormal{1/2}$ is the half-light radius. 
\newline\indent
Similar extended structures have also been discovered around Boötes I \citep{filionwyse2021, longeard2022, waller2023}, Ursa Major I \citep{waller2023}, Coma Berenices \citep{waller2023}, Ursa Minor \citep{sestito2023}, Fornax \citep{yang2022}, Hercules \citep{longeard2023}, and Sculptor \citep{sestito2023b}. One possible explanation for the presence of such anisotropic stars is dwarf-dwarf tidal interactions prior to infall \citep{genina2022}. However, it is interesting to ask whether such extended light profiles and apparent `tidal distortions' can occur via other means for tidally isolated systems. For example, \citet{tarumi2021} argue that galaxy mergers could explain the extended stellar halo around Tucana II.
\newline\indent
\citet{deason2022} discerned how different galaxy models affect the contribution of accreted stars to dwarf galaxy haloes, finding that minor mergers hold the strongest clues for dwarf galaxy models. 

Before the aforementioned discoveries, the existence of stellar haloes at these low mass scales remained inconclusive since they are thought to relate to early mergers that are less common as one goes down the mass scale of galaxies \citep{deason2022}. The mere existence of stellar haloes around dwarf galaxies helps to constrain the nature of galaxy formation and dark matter on the lowest mass scales.
\newline\indent
In this paper, we study the ellipticity and extended light around tidally isolated UFDs drawn from the `Engineering Dwarfs at Galaxy Formation’s Edge (EDGE) project \citep{agertz2020}. These tidally isolated galaxies are found to possess anisotropic and extended stellar outskirts, resembling the structure of tidal tails. The existence of these stellar haloes in the EDGE UFDs prompts us to look into their morphological origins, given that we intrinsically rule out the possibility of tidal isolation. Furthermore, we provide a comparison between the morphologies of the full suite of the EDGE simulations and an observed sample of UFDs, we use a Maximum Likelihood technique that was developed to calculate the observed structural parameters of dwarf galaxies \citep{martin2008, martin2016}. If the projected ellipticities for the entire EDGE simulation suite reasonably match observed dwarfs, then given that the EDGE UFDs are designed to be isolated from other massive systems, it is possible that tidal features in local UFDs are not necessarily due to tides and instead originate from a non-tidal scenario.
\newline\indent
This paper is organised as follows. In Section \ref{Methodology}, we discuss the setup of the EDGE simulations, the prerequisites we place on the selection of observed data samples, as well as the methods we use to derive the EDGE structural parameters in the fashion of an observational astronomer. In Section \ref{Results}, we describe the results obtained pertaining to the projected ellipticities of the EDGE simulations; the relationship between ellipticity and formation time, their comparison to observations,  as well as the extended stellar light of the faintest galaxies. We then examine these results and look towards their implications, discussing the origin of stellar ellipticity in the EDGE UFDs. Finally, we draw conclusions in Section \ref{Conclusion}.

\section{Method}\label{Methodology}
\subsection{EDGE Simulations} \label{EDGE section}
The suite of simulations examined in this work belongs to the EDGE project. These simulations were analysed using the {\sc tangos} database package \citep{pontzen2018} and the {\sc pynbody} analysis package \citep{pontzen2013}. A more comprehensive review of the simulations, and their underlying sub-grid physics, is found in \citet{agertz2020}.

The EDGE project is designed to study isolated UFDs with halo mass $10^9 < M/M_{\odot} < 5 \times 10^9$, in a simulated 50 Mpc void region. The simulations are initialised to assume cosmological parameters $\Omega_\textnormal{m} = 0.309$, $\Omega_\mathrm{\Lambda} = 0.691$, $\Omega_\mathrm{b} = 0.045$ and $H_0 = 67.77$ $\mathrm{km}$ $\mathrm{s^{-1}}$ $\mathrm{Mpc^{-1}}$, taken from the {\sc PLANCK} satellite 2013 data release \citep{planck2014}. The volume is initially simulated at a $512^3$ resolution, from $z = 99$ to $z = 0$. The largest void volume in this region is then selected and resimulated to a resolution of $2048^3$, with the inclusion of an appropriate small scale power to the grid. Within this resimulated region, the HOP halo Finder \citep{1998ApJ...498..137E} is implemented to find dark matter haloes at $z = 0$. Once a suitably isolated candidate has been confirmed, the halo is resimulated via the implementation of a zoom-in simulation technique \citep{katz1993, onorbe2014zoom}, up to redshift, $z=0$. 

We approach a maximum spatial resolution of $\sim$3 pc in the hydrodynamic grid. This high spatial resolution allows for the accurate injection of energy from a supernova, thus reducing the need for mechanisms required to prevent over-cooling of the supernovae-heated gas (see e.g \citealp{agertz2013, agertz2020, kimm2014, wheeler2019}).

The adaptive mesh refinement hydrodynamics code, {\sc ramses} \citep{teyssier2002}, is used to model the evolution of both baryonic matter and dark matter. The baryonic physics model makes use of a Schmidt law \citep{schmidt1959} to describe star formation in cells of gas that satisfy the required temperature and density (see \citealp{agertz2020}). Initially, each stellar particle represents $300 M_{\odot}$ and can be thought of as a mono-age stellar population described by a Chabrier initial mass function \citep{chabrier2003}.

The epoch of reionisation is modelled as a time-dependent uniform UV background at $z = 8.5$ \citep{haardt1996}. The reader may refer to \citet{rey2020} for details on the specific implementation of this model.

\subsection{Selection of observed candidates} \label{selection}
To provide a clear comparison of the shape distribution of observed dwarfs to the shape distribution of the EDGE simulation suite, we collate a refined list of dwarf galaxies belonging to the MW, presented in Table \ref{Table1}. We place two stringent constraints when creating this sample of galaxies. The first constraint is related to the mass of the galaxy, where we only include observed dwarfs that have stellar masses within approximately one order of magnitude of the EDGE dwarfs. For the second constraint, we adopt a list of observed galaxies that are thought to be tidally undisturbed (Kim et al., in preparation), since the EDGE galaxies were specifically selected due to their isolation. This second requirement is possible to meet thanks to the latest orbits taken from Gaia EDR3 \citep{gaiaedr32021}, which \citet{pace2022} then combine with accurate photometry to determine the systemic proper motions, thus providing insight into their respective tidal interactions. It should be noted that we use the pericentre values including the influence of the LMC, to provide us with the most authentic orbital scenarios. 

The metric for isolation of an UFD is categorised as a ratio between its tidal and half-light radius. However, the time variation of the tidal radius and its dependency on the mass distribution of both systems involved are usually poorly understood; hence the tidal radius remains ambiguously defined \citep{simon2019}. A solution for this uncertainty is to approximate the tidal radius as the position where the total force (from the satellite and host) matches the centrifugal acceleration needed to stay on the same orbit as the satellite. The tidal radius is given as follows \citep{hoerner1957, binney2008, simon2019}: 

\begin{equation}
    r_{\textnormal{t}} = \left(\frac{m_{\mathrm{dwarf}}}{3M_{\textnormal{MW}}}\right)^{\frac{1}{3}} d,
    \label{eq1}
\end{equation}

\noindent where $r_\textnormal{t}$ is the tidal radius, $m_{\textnormal{dwarf}}$ is the dwarf galaxy mass, $M_{\textnormal{MW}}$ is the MW mass enclosed within the dwarf orbital radius, and $d$ is the distance between the dwarf and the galactic centre of the host system. It should be noted that Equation \ref{eq1} is only an approximation, as it assumes a point-mass approximation for the dwarf and the MW, purely radial motion between the dwarf and the MW, and stars within the dwarf moving on purely radial orbits \citep{read2006c}.

Following \citet{simon2019}, we define dwarfs to be “tidally isolated” if they have $\frac{r_\textnormal{t}}{r_\textnormal{1/2}}$ > 3. In practice, even these dwarfs are likely stripped to some degree. \citet{shipp2023} recently showed using the FIRE simulations that many "intact" looking satellites have tidal tails, with $\sim 66\%$ of a total 64 stream progenitors being mistaken for intact satellites. However, it will be challenging to detect any stripping from their extended light distributions. The only exception to this is dwarfs that interact with other dwarfs before infall to the MW. These can be on apparently benign orbits, where the galaxy shows no present-day sign of prior tidal interactions, despite having experienced significant stripping in the past \citep{genina2022}. At present, no method has been proposed for distinguishing such dwarfs from genuinely tidally isolated systems. As such, we must accept the possibility that some of our samples will be contaminated by such tidally affected systems. Nevertheless, in a $\Lambda$CDM cosmology, these are expected to be quite rare, with \citet{genina2022} stating that they find 9 out of 212 simulated luminous dwarfs as analogues of this scenario.

\begin{table*}
\centering
\renewcommand{\arraystretch}{1.4}
\begin{tabular}{cccccc}
    {Dwarf Galaxy} & {Ellipticity, $\varepsilon$} & {Pericentres (kpc)} & {Half-light radius, $r_{1/2}$ (kpc)} & {Tidal radius, $r_{\textnormal{t}}$ (kpc)} & {References$^a$} \\
    \hline
    Sculptor & 0.32 ± 0.03 & 44.9 $^{+4.3}_{-3.9}$ & 0.279 ± 0.016 & 1.050 & (1, 12)\\
    Leo I & 0.21 ± 0.03 & 47.5 $^{+30.9}_{-24.0}$ & 0.270 $^{+0.017}_{-0.016}$ & 1.082 & (1, 12)\\
    Leo II & 0.13 ± 0.05 & 61.4 $^{+62.3}_{-34.7}$ & 0.171 ± 0.010 & 0.970 & (1, 12)\\
    Ursa Minor & 0.56 ± 0.05 & 55.7 $^{+8.4}_{-7.0}$ & 0.405 ± 0.021 & 1.421 & (1, 12)\\
    Sextans & 0.35 ± 0.05 & 82.2 $^{+3.8}_{-4.3}$ & 0.456 ± 0.015 & 1.750 & (1, 12)\\
    Carina & 0.33 ± 0.05 & 77.9 $^{+24.1}_{-17.9}$ & 0.311 ± 0.015 & 1.312 & (1, 12)\\
    Canes Venatici I & 0.39 ± 0.03 & 84.5 $^{+53.6}_{-37.2}$ & 0.437 ± 0.018 & 1.717 & (2, 12)\\
    Canes Venatici II & 0.52 $^{+0.10}_{-0.11}$ & 47.4 $^{+46.8}_{-29.7}$ & 0.071 ± 0.011 & 0.437 & (2, 12)\\
    Draco & 0.31 ± 0.02 & 58.0 $^{+11.4}_{-9.5}$ & 0.231 ± 0.017 & 1.180 & (2, 12)\\
    Ursa Major I & 0.80 ± 0.04 & 49.9 $^{+46.2}_{-15.6}$ & 0.295 ± 0.028 & 0.962 & (2, 12)\\
    Leo T & 0.29 $^{+0.12}_{-0.14}$ & > 250 & 0.118 ± 0.011 & - & (2)\\
    Leo IV & 0.49 ± 0.11 & 66.8 $^{+60.7}_{-44.1}$ & 0.114 ± 0.013 & 0.527 & (3, 12)\\
    Leo V & 0.50 ± 0.15 & 165.8 $^{+5.8}_{-49.2}$ & 0.049 ± 0.016 & 0.636 & (3, 12)\\
    Leo P & 0.52 & - & - & - & (4)\\
    Reticulum II & 0.60 ± 0.10 & 37.0 $^{+2.9}_{-5.3}$ & 0.051 ± 0.003 & 0.262 & (5, 12)\\
    Pisces II & 0.40 ± 0.10 & 130.5 $^{+70.1}_{-72.3}$ & 0.060 ± 0.010 & 0.991 & (6, 12)\\
    Eridanus II & 0.48 ± 0.04 & 114.4 $^{+80.9}_{-67.6}$ & 0.246 ± 0.017 & 1.683 & (7, 12)\\
    Bo\"otes I & 0.68 ± 0.15 & 37.9 $^{+7.5}_{-6.8}$ & 0.191 ± 0.008 & 0.517 & (8, 12)\\
    Hydrus I & 0.21 $^{+0.15}_{-0.07}$ & 45.8 $^{+16.1}_{-6.0}$ & 0.053 ± 0.004 & 0.270 & (9, 12)\\
    Pegasus III & 0.46 $^{+0.18}_{-0.27}$ & 141.0 $^{+87.8}_{-79.3}$ & 0.078 $^{+0.031}_{-0.025}$ & 1.151 & (10, 12)\\
    Hercules & 0.67 ± 0.03 & 67.4 $^{+15.5}_{-16.1}$ & 0.216 ± 0.020 & 0.877 & (11, 12)\\
\end{tabular}
    \caption{\label{Table1} Observed ellipticities, pericentres, half-light radii and calculated tidal radii for 21 dwarf galaxies in our LG. 
    $^a$References: (1) \citet{irwin1995}, (2) \citet{martin2008}, (3) \citet{dejong2010}, (4) \citet{leop2015}, (5) \citet{rettwo}, (6) \citet{belokurov2010}, (7) \citet{eritwo2016}, (8) \citet{pristine2021}, (9) \citet{mcconachie2006}, (10) \citet{kim2015a}, (11) \citet{sand2009}, (12) \citet{pace2022}.}
\end{table*}

\subsection{Structural parameters} \label{structural params}
To derive the structural parameters of the EDGE simulations, we employ a Maximum Likelihood technique similar to the one utilised by observational astronomers to uncover the structural parameters for dwarf galaxies in the SDSS \citep{martin2008} and PandAS \citep{martin2016} surveys\footnote{PandAS is an astronomical survey focused on the content and structure of M31 and M33 \citep{pandas2014}.}. In this paper, the method is used in such a way as to treat the simulations as if they were two-dimensional observations projected onto the sky, so that the resemblance between the EDGE simulations and observational data can be accurately analysed.

Equation \ref{likeeq} calculates the probability each star particle contributes to the structural parameters, given its respective position. As the stellar particles in EDGE are representative of a stellar mass on the order of $10^2 M_{\odot}$, we can determine how much each stellar particle should contribute to the calculation of the structural parameters by weighting via the stellar mass.

\begin{equation}\label{likeeq}
    l_i = \frac{1.68^{2}N_{*}}{2\pi  r_{\textnormal{1/2}}^{2} (1-\varepsilon)} \exp\left(\frac{-1.68r_i}{r_{\textnormal{1/2}}}\right) \frac{m_{i}}{M_*},
\end{equation}
\\
\noindent where the relation between the half-light radius and the exponential scale radius of the profile is $r_\textnormal{1/2} \approx 1.68r_\textnormal{e}$, $N_*$ is the number of stars in the sample, $M_*$ is the total stellar mass, $m_i$ is individual stellar mass, $\varepsilon$ is the ellipticity defined as $\varepsilon = 1 - b/a$, with $b/a$ as the minor-to-major-axis ratio of the system, $\theta$ is the position angle of the major axis, defined as East of North, $r_\textnormal{1/2}$ is the half-light radius of its assumed exponential radial profile, and $r_i$ is the elliptical radius. Here, $r_i$, is related to the spatial positions $x_i$ and $y_i$ as follows, 

\begin{equation}
    r_i = \left(\left(\frac{1}{1-\varepsilon}(x_{i}\cos\theta - y_{i}\sin\theta)\right)^{2}+ \left(x_{i}\sin\theta + y_{i}\cos\theta\right)^{2}\right)^{\frac{1}{2}}
\end{equation}
\\
A more comprehensive mathematical approach may be found in \citet{martin2008}, where a similar exponential model is used to describe a low stellar density. 

The total log-likelihood is calculated by taking the summation of all the logged individual probabilities, 

\begin{equation}
    \log\mathcal{L} = \sum_{i} \log l_i
\end{equation}
\\
We determine the most likely shape parameters $(\varepsilon, \theta, r_\textnormal{1/2})$ for each EDGE UFD with the {\sc emcee} code \citep{foreman2013}. This is a Markov Chain Monte Carlo (MCMC) method, and it is found to definitively converge upon the structural parameters of the galaxy when 50 walkers are run over a total of 600 steps. Similar to \citet{martin2016}, we place flat priors for the three parameters such that $0 \leqslant \varepsilon < 1$, $\theta$ is in an interval of 180 degrees, and $r_\textnormal{1/2} > 0$.

\subsection{Cutting on surface brightness} \label{SB cut}
Since one of our main aims is to compare simulations to observations, we attempt to replicate the same methods and techniques that observational astronomers use. Therefore, during the MCMC calculation we apply a surface brightness cut to the EDGE simulations. This creates the effect that our mock observations of simulations are limited by surface brightness, just as observations are through the use of telescopes. 

Inspired by the literature, we place two different surface brightness cuts on the simulated UFDs. The first cut is at 25.5 mag arcsec$^{-2}$ since this was the surface brightness limit of the SDSS telescope \citep{whiting2007, koposov2008} used to observe several MW dwarf galaxies in Table \ref{Table1}. We place the next surface brightness cut at 30 mag arcsec$^{-2}$ to highlight what we should be able to predict with the most modern detection instruments. The latter cut is a more optimistic approach inspired by the contemporary advances of observational astrophysics in recent years, i.e. the Dark Energy Survey (DES) \citep{des2021}, the DECam Local Volume Exploration survey (DELVE) \citep{dw2021}, and within the next few years, the Vera C. Rubin Observatory \citep{lsst2018}.

\subsection{Gaussian KDE fitting method} \label{kde}
The ellipticity we calculate with our Maximum Likelihood technique is the projected ellipticity of the UFD and not the true ellipticity. Accordingly, we orient each EDGE galaxy around 100 random viewing angles to create a probability distribution function (hereafter PDF) of projected ellipticities. 

All our PDFs are created with a kernel density estimator (KDE), used to convolve our data with a Gaussian kernel. A certain degree of smoothing is employed to produce the PDFs of projected ellipticity for both EDGE and observations. We use Silverman's Rule to define the degree of this smoothing so that the PDF provides a match to the underlying data. With each PDF, we include a rug plot displaying the data points from which the distributions are constructed.

\section{Results}\label{Results}
The {\sc GenetIC} code \citep{stopyra2021} is utilised within the EDGE simulations to `genetically modify' the initial conditions for an EDGE UFD ($t_{\mathrm{form}} = 2.4$ Gyrs) to form three unique variations at earlier ($t_{\mathrm{form}} = 2.8$ Gyrs) and later times ($t_{\mathrm{form}} =$ 3.1 Gyrs, 3.6 Gyrs) \citep{rey2019}. We define formation time as the time when the galaxy has assembled 50\% of its final mass at $z=0$.

\citet{rey2019} studied the mass accretion histories for these UFDs, and revealed that the later-forming variations assemble their stellar mass from late-time dry mergers. Such an assembly history leads to extremely low surface brightness and an increase in the half-light radius (i.e. an increase in the size of the galaxy). Conversely, the mass accretion history of the earlier-forming galaxy primarily consisted of stars that had formed in-situ, leading to a higher surface brightness and a decrease in the half-light radius.

\begin{figure}
    \includegraphics[width=\columnwidth]{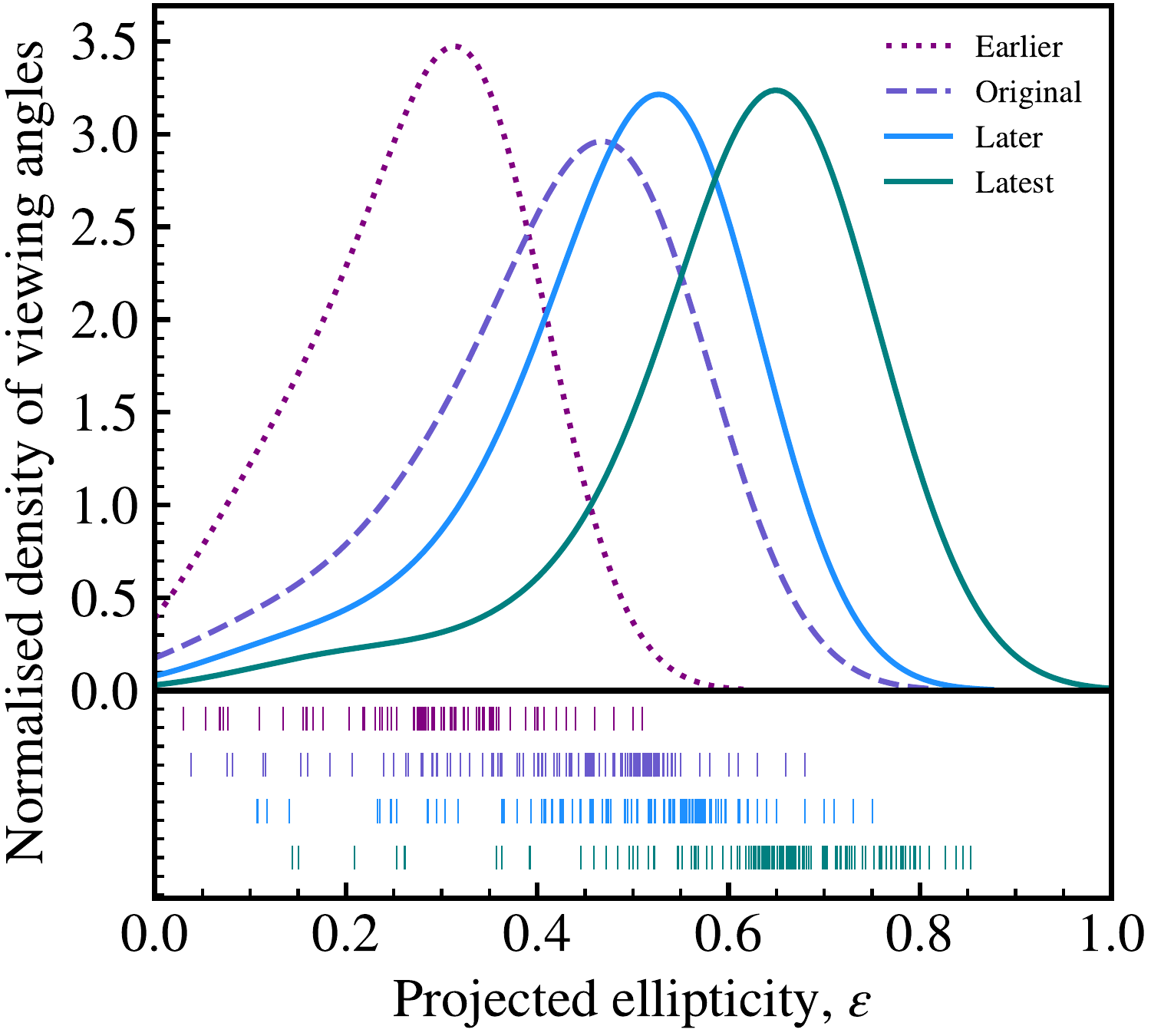}
    \caption{PDFs of projected ellipticity for a fiducial EDGE UFD (original - dashed) and three variations of this simulated UFD, genetically modified to have formation times; earlier (dotted), later (blue), and latest (green). These four distributions of projected ellipticities are created at a surface brightness cut of 30 mag arcsec$^{-2}$. The coloured lines in the underlying rug plot represent the individual projected ellipticities in the respective distribution.}
    \label{1459all}
\end{figure}

\begin{figure}
    \includegraphics[width=\columnwidth]{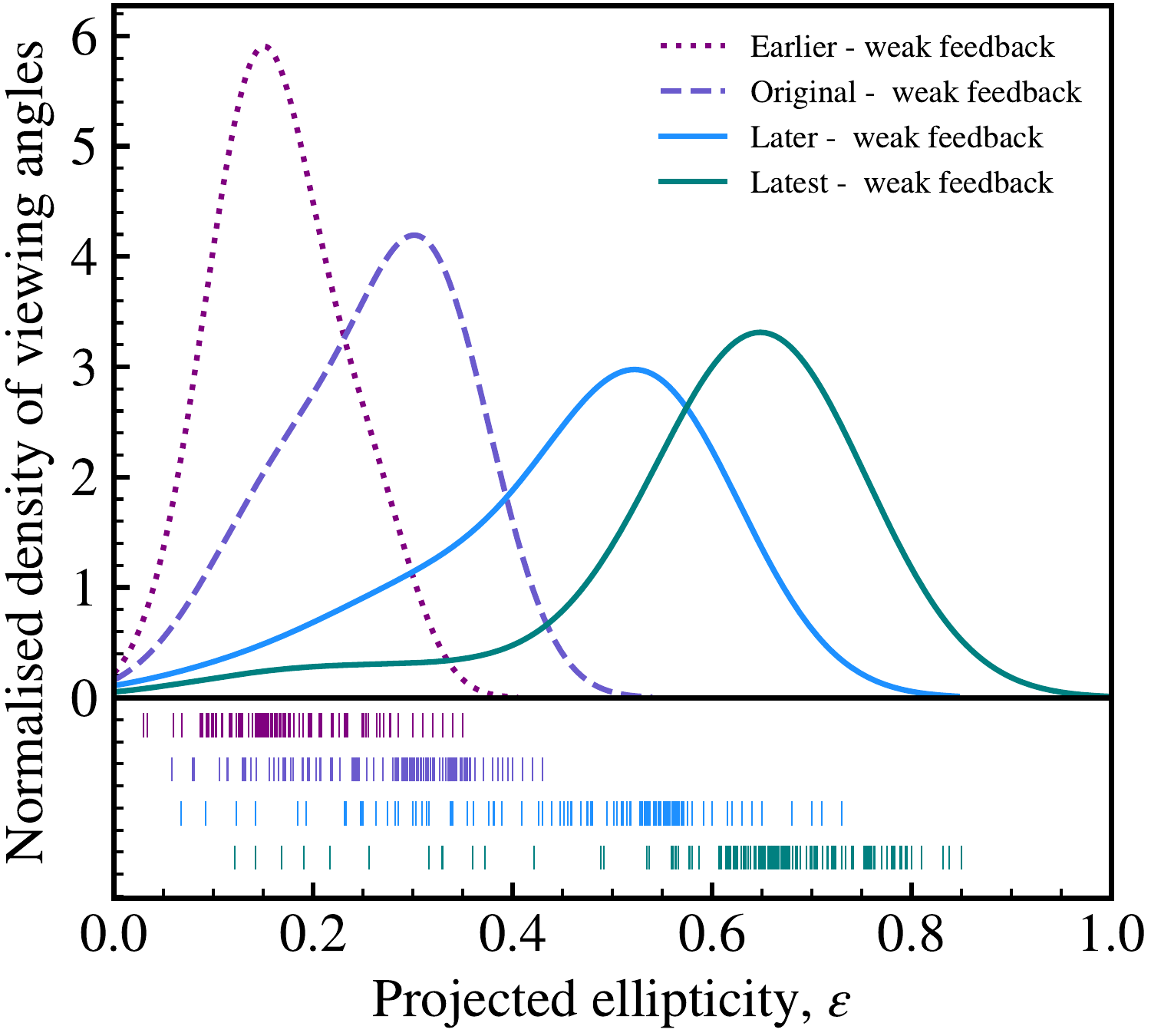}
    \caption{PDFs of projected ellipticity for a fiducial EDGE UFD created in an environment with limited feedback (original - purple dashed) and three variations of this simulated UFD, genetically modified to have formation times; earlier (pink dotted), later (blue), and latest (green), also created with the `weak feedback' model. Once again, these four distributions of projected ellipticities are created at a surface brightness cut of 30 mag arcsec$^{-2}$.The coloured lines in the underlying rug plot represent the individual projected ellipticities in the respective distribution.}
    \label{1459all_fblim}
\end{figure}

\subsection{EDGE projected ellipticity correlates with formation time} \label{time vs shape}
Figure \ref{1459all} displays the PDFs of projected ellipticity for the same fiducial UFD and three variations of this fiducial UFD at earlier and later times. The projected ellipticities here are taken at a surface brightness cut of 30 mag arcsec$^{-2}$.

A systematic shift from a lower projected ellipticity to a higher projected ellipticity is seen from the peak of the distributions. This shift in projected ellipticity correlates with the formation time of the UFDs, i.e. earlier assembly times have lower ellipticities, and later assembly times have greater ellipticities. Our findings represent the first evidence in support of a causal relationship between the time of main halo formation in an UFD and the galactic stellar ellipticity of an UFD. 

To provide complete clarity, the ellipticity of the stellar content is directly related to the distribution of the stellar structure \citep{rey2019}. The distribution of the stellar content in UFDs will depend on whether star-forming gas is available. If it is, the majority of the UFD forms via in-situ star formation, and this leads to a rounder, more compact shape. However, if the gas escapes the UFD, no more star formation will occur, and the UFD will form via ex-situ mergers, leading to a fainter, more elliptical shape. Therefore, the ellipticity is determined by the method of formation of the stellar component, which is decided by the availability of gas in the galaxy. However, this availability of gas is decided by the formation time of the main halo and whether it is large enough at reionisation to retain its gas. And so, the formation time of the main halo dictates the availability of gas, which in turn decides whether the stars originate in-situ or ex-situ, i.e. where the ellipticity arises from.

Furthermore, we must now consider that this relationship is affected by how much stellar mass forms in-situ in the centre of the UFD, and this becomes very sensitive to the choice of feedback model. A model with less feedback will produce a lot more in-situ stars, as less feedback implies an increase in star-forming gas. There will also be an increase in the late-time accreted component of the stellar content, however, it becomes uncertain how our relationship will now follow with ellipticity.

Figure \ref{1459all_fblim} displays the PDFs of projected ellipticity at a surface brightness cut of 30 mag arcsec$^{-2}$ for the four variations of the UFD in Figure \ref{1459all}, now implemented with a `weak-feedback' model. This model artificially limits the efficiency of the supernovae wind driving by placing numerical limits on the maximum supernovae gas temperatures and velocities (see \citealp{agertz2013, agertz2020}).

Once again, we see a clear systematic shift from a lower projected ellipticity to a higher projected ellipticity from the peaks of the distributions. This shift in projected ellipticity correlates with the formation time of the UFDs going from lower ellipticities at earlier assembly times to higher ellipticities at later assembly times. Therefore, even if our feedback model does not provide an absolute description of reality, under the regime of this `weak feedback' model, the relative ordering of ellipticity with mass growth histories stands. Our relationship is robust to the feedback physics we use, and the stars at large radii that make up these extended tails are found within two different feedback models, making them a strong prediction from our simulations.

We observe that the PDFs shown in Figure \ref{1459all} have a great deal of overlap, calculating a 12\% chance to mistake the projected ellipticity of the earlier-forming UFD with the latest-forming UFD. Meanwhile, there is only a 2\% chance to mistake the projected ellipticity of the earlier-forming UFD with the latest-forming UFD, from the PDFs shown in Figure \ref{1459all_fblim}.

\begin{figure} 
    \includegraphics[width=\columnwidth]{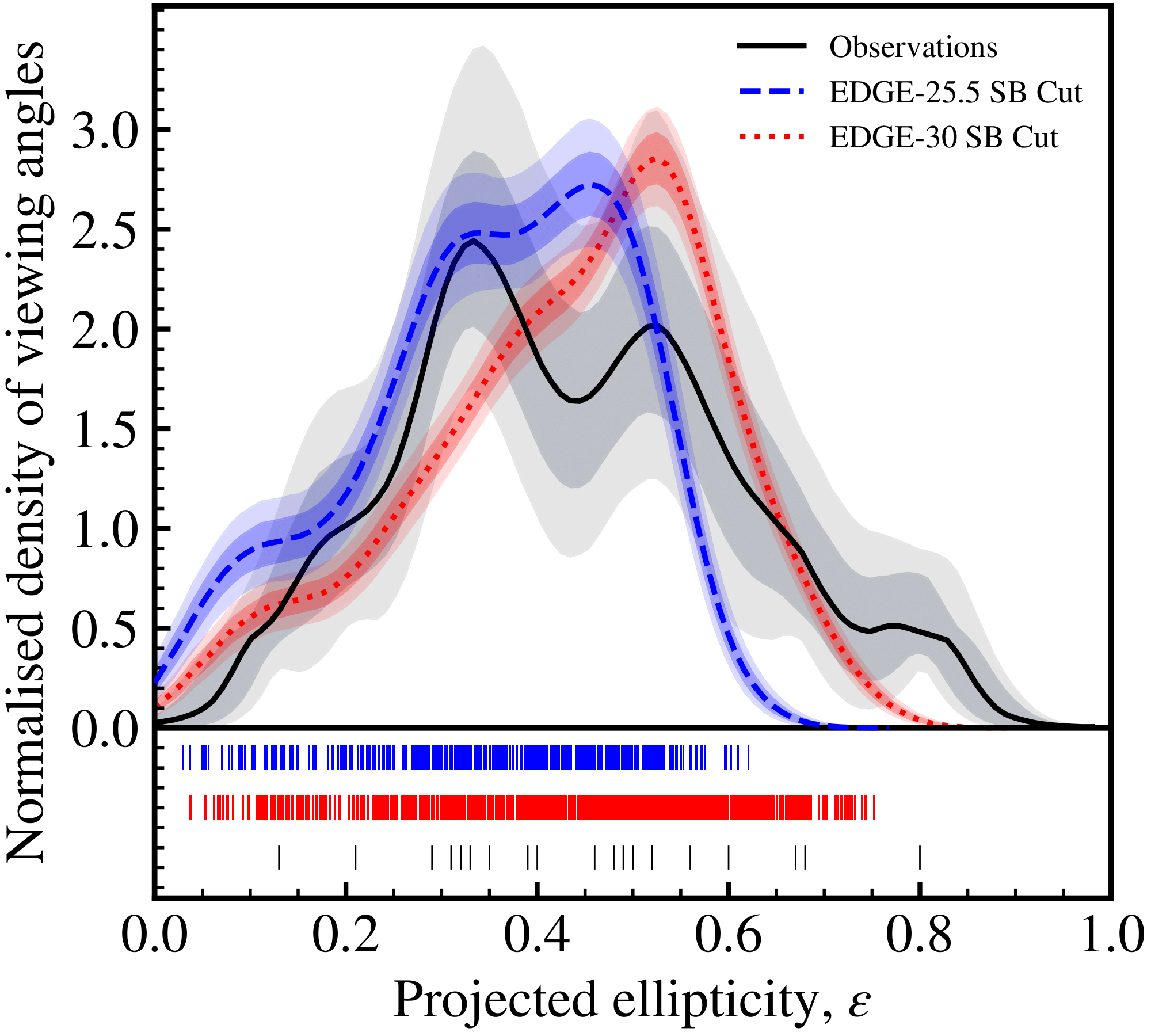}
    \caption{PDFs of projected ellipticity from the sample of observed dwarf galaxies in Table \ref{Table1} (black/grey) and the EDGE simulation suite (blue/red). The median, 1$\sigma$, and 2$\sigma$ confidence intervals on the observed distribution are calculated from the uncertainties of the ellipticities in Table \ref{Table1}. There are two PDFs for the EDGE simulations, the blue dashed line represents the distribution of projected ellipticities of the EDGE simulations when observed up to a surface brightness limit of 25.5 mag arcsec$^{-2}$, and the red dotted line represents the distribution of projected ellipticities of the EDGE simulations when observed up to a surface brightness limit of 30 mag arcsec$^{-2}$. The 1$\sigma$ and 2$\sigma$ confidence intervals on the EDGE data result from a bootstrapping technique used to determine the variation in the EDGE ellipticity distributions, which are randomly sampled from the weighting by formation time. In the legend, the acronym `SB' denotes surface brightness. The coloured lines in the underlying rug plot represent the individual projected ellipticities in the respective distribution.}
    \label{EDGEOBS}
\end{figure}

\subsection{Projected ellipticity in EDGE vs observed dwarf galaxies} \label{ecomparison}
The PDFs of the projected ellipticities for the EDGE simulation suite are shown in Figure \ref{EDGEOBS}. Here, we use the full suite of the EDGE simulations to make our comparison with observations. To see a tabulated summary of the 10 EDGE UFDs, please refer to \citet{rey_edge_2021}. We note that only five of these ten UFDs are unique realisations; we construct the additional five UFDs from genetic modifications applied to the original five.
\newline\indent
The EDGE PDFs are given at surface brightness cuts of 25.5 mag arcsec$^{-2}$ (blue dashed) and 30 mag arcsec$^{-2}$ (red dotted). Also shown in Figure \ref{EDGEOBS} is the PDF of the projected ellipticities created from the observed ellipticity samples (black) in Table \ref{Table1}, along with the respective confidence intervals at 68\% (1$\sigma$) and 95\% (2$\sigma$) variance, calculated from the uncertainties of the observed samples.
\newline\indent
From the newly discovered relationship between an UFDs formation time and ellipticity, we know that our more elliptical galaxies are later forming and the less elliptical galaxies are earlier forming. Studying the populous of haloes in the EDGE volume, it becomes clear that the haloes forming earlier and later are significantly rarer. Therefore, when choosing to weight our EDGE PDFs of projected ellipticity, we do so via their formation times opposed to stellar mass. We assume that formation time is the dominant variable beyond mass when it comes to the ellipticity of these galaxies. To weight the EDGE PDFs by formation time, we use our probability density function of haloes in the EDGE volume to weight the contribution of each EDGE UFD to the shape distribution. We note that we only include the formation times of systems with similarly sized halo masses to those of our EDGE samples. This weighting ensures that the EDGE PDFs are not biased by formation time. The UFDs with a more common formation time will contribute more to the EDGE PDFs, and the UFDs with a rarer formation time will contribute less to the EDGE PDFs, thus creating a more accurate comparison to the PDF from our sample of observed dwarfs.
\newline\indent
The comparison between EDGE PDFs presented in Figure \ref{EDGEOBS} displays the peak of the EDGE distribution shifting towards higher ellipticities at a fainter surface brightness cut of 30 mag arcsec$^{-2}$. As these galaxies are viewed out to a fainter surface brightness limit, the number of stars increases in our `observations' of the EDGE UFDs. Therefore, the increase in projected ellipticity confirms that these galaxies have an elongated stellar distribution along one axis. 
\newline\indent
It is important to highlight the inhomogeneity of surface brightness limits the galaxies in our sample were observed at. A number of ellipticity measurements were taken at a surface brightness limit $\sim$ 25.5 mag arcsec$^{-2}$, but for some galaxies observed more recently, the surface brightness limits extend up to $\sim$ 30 mag arcsec$^{-2}$, thanks to improved detection instruments. Following the above-mentioned relation in EDGE, if some galaxies were observed out to a fainter surface brightness limit, they would sit at higher ellipticities on the observed PDF, and this could explain the existence of the second smaller peak in the observed distribution.
\newline\indent
The projected ellipticities of EDGE galaxies viewed at a more luminous surface brightness limit of 25.5 mag arcsec$^{-2}$ provide good agreement with a significant number of dwarf galaxies from the observed sample, with one of the peaks of the blue EDGE distribution laying well within the overlap of the larger observed peak. We also see a good agreement between the smaller peak of observed galaxies and the fainter distribution of EDGE galaxies at 30 mag arcsec$^{-2}$. These agreements point towards a relation between the shape of the EDGE UFDs and the sample of observed dwarfs up to a surface brightness limit of 30 mag arcsec$^{-2}$. As our EDGE UFDs are tidally isolated, this suggests a number of the galaxies in our sample of observed dwarfs have ellipticities that are perhaps unduly attributed to tides. This possible tidal isolation is further reinforced by the tidal radii we calculate from the large pericentres of the observed sample.
 
\subsubsection{A cautionary note on tidal effects in dwarf galaxies}
Following our discussion in Section \ref{selection}, even though we make an effort to try to reduce the number of tidally affected galaxies in Table \ref{Table1}, there is still the possibility that tides may have some influence on observations, and even these dwarfs in our sample are likely stripped to some degree. For example, \citet{shipp2023} report from their simulations that more than 50\% of satellites have tidal tails at distances 50-200 kpc from their hosts. Consequently, this may be the reason why we see the PDF of observed galaxies skew to higher projected ellipticities in Figure \ref{EDGEOBS}, where we notice from the underlying rug plot that there is a handful of galaxies lying beyond the second peak in the observed distribution.

Alternatively, we also have to account that the present-day orbit does not always provide the complete history of a galaxy. Given the density of the environment that these dwarfs live in, it is plausible they experienced previous interactions with close-by dwarf galaxies such as the LMC or even with each other, which may yet be another reason explaining the high ellipticity population in our observed sample. Using the example of Fornax, \citet{genina2022} show that systems appearing to be tidally isolated today can have had significant galaxy-galaxy interactions in the past. Analogues of this scenario appearing in our observed sample would shift the distribution of observed galaxies to a more elliptical peak. However, as previously stated in Section \ref{selection}, this scenario is expected to be quite rare, with \citet{genina2022} finding a <5\% chance of a Fornax analogue from their sample of 212 simulated dwarfs.

\begin{figure*}
    \includegraphics[width=17.5cm]{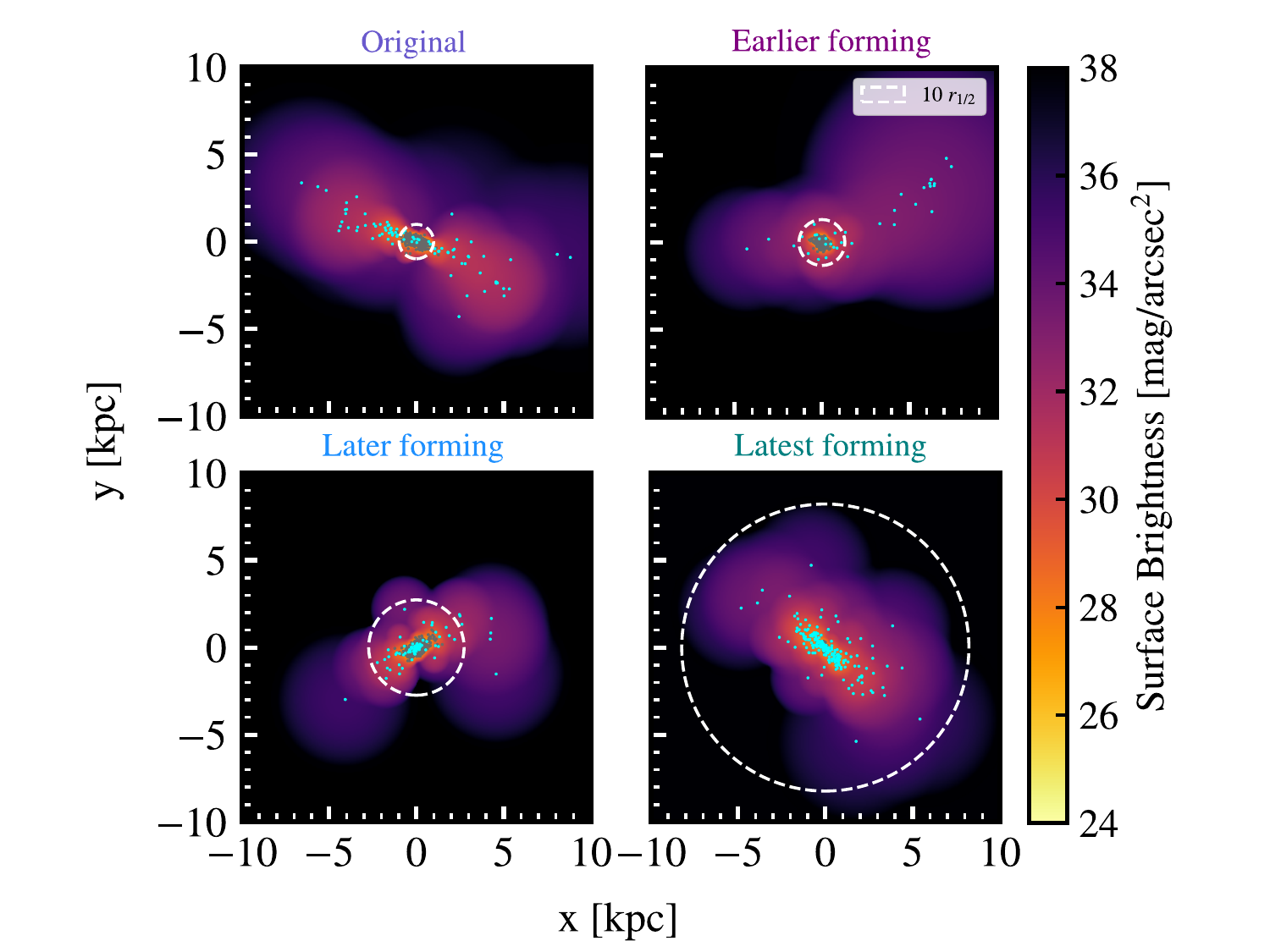}
    \caption{Spatial distributions of the stellar content in four EDGE UFDs, with underlying surface brightness maps created via a kernel smoothing scheme. The projection axes of the galaxies are oriented in such a way as to maximise the ellipticity of the stellar distribution. Top left panel shows the spatial distribution of the original fiducial simulation. Top right panel shows the spatial distribution of the original UFD, genetically modified to form at an earlier time. Bottom left panel shows the spatial distribution of the original UFD, genetically modified to form at a later time. Bottom right panel shows the spatial distribution of the original UFD, genetically modified to form at the latest time of the four. Notice that in each case, the extended stellar haloes are highly anisotropic. These look like tidal tails but owe instead to the late-time accretion of lower-mass companions. Star particles that stem from this accretion origin are coloured cyan. Meanwhile, star particles coloured grey represent stellar content that formed in-situ. The radial distance of the white dashed circle represents ten projected half-light radii.}
    \label{mughsots}
\end{figure*}

\subsection{Origin of extended stellar content in EDGE}
Figure \ref{mughsots} represents the present-day spatial distributions for the same UFD in Figure \ref{1459all}, along with its three variations with altered formation times, and their underlying surface brightness maps. The surface brightness maps were created with a kernel smoothing scheme using the {\sc py-sphviewer} package \citet{alejandro_benitez_llambay_2015_21703}. 
\newline\indent
As these are projected spatial distributions, they are susceptible to the two-dimensional effect of appearing seemingly less elliptical than their true galactic shapes imply. To counter this, we ensured that the stellar contents of all the galaxies were observed from their most elliptical viewing angle.
\newline\indent
The first key result that stands out when observing the stellar content in Figure \ref{mughsots} is the elongation that the systems possess along one axis. This elongation coincides with the stars coloured cyan, which represent stars that come from a late-time dry accretion origin. The stars coloured grey, on the other hand, are stars that form from an in-situ origin and we see these are much more central in the stellar distribution.
\newline\indent
While the UFDs in the bottom two panels of Figure \ref{mughsots} are the most elliptical in terms of the overall shape of the galactic centre, the extended stellar distributions of the top two panels are still very clearly discerned upon inspection.
\newline\indent
The centres of the two UFDs residing in the top panels of Figure \ref{mughsots} are not as elliptical, however, these galaxies still possess this extended distribution feature, with detritus of member stars located up to $\sim$ 78 half-light radii from the centre of the original galaxy and $\sim$ 56 half-light radii from the centre of the earlier-forming galaxy. To provide a visual reference for these large half-light radii, we include a white dashed circle in our spatial distributions, for which the radial distance from the centre represents ten half-light radii. From the surface brightness maps, we see that these outer stars at these distances will become observable at a surface brightness $\gtrsim$ 34 mag arcsec$^{-2}$. We note how great the radial distance to ten half-light radii is for the later-forming galaxies, implying how faint they are compared to the earlier-forming UFDs.
\newline\indent
The original UFD in the upper left panel of Figure \ref{mughsots} has an apparent elongated distribution of stars and possesses a much larger spread of stars than either of the later-forming galaxies. This original galaxy has an ex-situ mass fraction of 2.96\%. 

The earlier-forming UFD in the upper right panel is much rounder at the centre, which is supported by the low projected ellipticity distribution shown in Figure \ref{1459all}. The earlier-forming galaxy has an ex-situ mass fraction of 1.05\%, making it the UFD with the lowest amount of ex-situ content compared to the other three. 

The later-forming UFD in the bottom left panel has less of an extended distribution than the original and early-forming UFDs, however, its galactic centre is more elliptical. The later-forming galaxy has an ex-situ mass fraction of 7.31\%. 

Finally, the latest-forming UFD in the bottom right panel is strongly elongated, with the most elliptical centre out of the four UFDs. The latest-forming UFD has the highest content of ex-situ stars with a mass fraction of 95.8\%, given that practically all stellar content originates from ex-situ dry accretions. Also, the overall stellar mass for this galaxy is an order of magnitude smaller than the other three UFDs. Justified from the peaks of the PDFs in Figure \ref{1459all}, we can definitively say that the latest-forming UFD has the most elliptical galactic centre. These PDFs consist of ellipticities taken at a surface brightness cut of 30 mag arcsec$^{-2}$, meaning they exclude the majority of the extended stellar outskirts.

We trace back the origin of the stars in the EDGE UFDs and find that those coloured cyan in Figure \ref{mughsots} accrete onto the systems through late-time dry mergers and make up the extended part of the stellar distribution. The stellar ellipticities in the EDGE UFDs thus originate from these late-time accretion events of smaller haloes that had their gas quenched by reionisation before they could merge onto the main halo. The bottom two panels of Figure \ref{mughsots}, showing the later-forming galaxies, exemplify this as they assemble primarily through late-time dry accretion and have extended distributions of stars along one axis around the galactic centres, giving them systematically larger ellipticities. 

On the other hand, the earlier-forming UFDs form their stellar content in-situ and have less elliptical galactic centres, with systematically lower ellipticities. However, in the spatial distributions of the two earlier-forming galaxies, there is still a distribution of stellar material coloured cyan, extending even further than the elongations depicted in the later-forming galaxies. Therefore, we show that even though these galaxies formed earlier, they still have elongated stellar content originating from the late-time accretion events of smaller haloes. Such features clarify that the extended starlight in EDGE is a natal characteristic of these ultra-faint systems, and these cyan coloured stars from late-time dry accretion events are the origin of this stellar ellipticity. For the first time, we show that the accretion mechanism giving rise to the extended shapes emerges naturally in a fully cosmological context, where several dark matter sub-haloes undergo hierarchical minor mergers following $\Lambda$CDM cosmology. And the existence of these stellar haloes around the smallest galaxies in the Universe constrains $\Lambda$CDM at the smallest scales.

\section{Conclusions} \label{Conclusion}
Within our local volume of the Universe, we find a multitude of the faintest, most dark matter dominated galaxies. Given that these UFDs are scattered around more massive systems, it is logical to assume that these systems are extremely prone to tidal effects. Thus, the large projected ellipticities of the stellar distribution and the stellar detritus of these systems can easily be attributed to tidal deformation. 

Utilising the EDGE simulations, we have shown that despite their tidal isolation, our simulated dwarfs exhibit anisotropic extended stellar outskirts that masquerade as tidal tails but are instead natal, owing to the origin of a late-time dry accretion assembly. Furthermore, we revealed that UFDs with later formation times have more elliptical stellar distributions, thus establishing a novel connection between the shape of an UFD and its respective formation time. This newly discovered relationship was robust to a wide variation in the feedback model, making it a strong prediction from our simulations. The above-mentioned results extend the conclusion found in \citet{rey2019}, in which the authors discovered that UFDs with later formation times have an extremely low surface brightness and a much larger stellar size. These discoveries within the EDGE fully cosmological context are vital in assessing the extent to which the smallest galaxies in the Universe have mechanisms and features indistinguishable from more massive galaxies, such as the MW, with respective fossil records detailing their history.

We studied the projected stellar ellipticities of 10 isolated UFDs in the EDGE cosmological simulation suite by implementing a well-known observational method to calculate structural parameters. The Maximum Likelihood technique, developed and employed by observational astronomers \citep{martin2008, martin2016}, allowed us to make direct comparisons concerning the projected ellipticity of our simulations, contrasted to a refined sample of observed MW dwarf galaxies in our LG.
\newline\indent
We sampled projected ellipticities from around 100 random viewing angles for each EDGE UFD to acquire a representative distribution of orientations. Observing the EDGE UFDs out to fainter surface brightness, we noticed an increase in projected ellipticity. Therefore, if UFDs have extended stellar distributions in reality, we should expect a similar increase in the projected ellipticity of known UFDs when these galaxies are further uncovered with deeper and better-resolved spectra of their surrounding stars.
\newline\indent
The PDFs of projected ellipticity for EDGE and observations displayed a good agreement, given that peaks of their distributions lay comfortably within the confidence intervals. This agreement implies our simulated EDGE UFDs resemble the shapes of a number of faint dwarfs belonging to the MW. As the EDGE UFDs are designed to be isolated from more massive systems, it is possible that tidal features in these observed dwarfs are not necessarily due to tides but instead originate from a non-tidal scenario. This theorised isolation of our select sample of MW dwarf galaxies is reinforced by their newfound orbital parameters in \citet{pace2022}.
\newline\indent
If a significant number of the nearby UFDs in our LG are tidally intact, as our results may suggest, their baryonic and dark matter contents could remain uninfluenced from more massive systems in their surrounding environments. Therefore, many LG UFDs would serve as excellent natural laboratories for probing dark matter and galaxy formation physics.

\section*{Acknowledgements}
This work was performed using the DiRAC Data Intensive service at Leicester, operated by the University of Leicester IT Services, which forms part of the STFC DiRAC HPC Facility (www.dirac.ac.uk). The equipment was funded by BEIS capital funding via STFC capital grants ST/K000373/1 and ST/R002363/1 and STFC DiRAC Operations grant ST/R001014/1. DiRAC is part of the National e-Infrastructure. OA and EA acknowledge financial support from the Knut and Alice Wallenberg Foundation and the Swedish Research Council (grant 2019-04659). EA also acknowledges support from the US NSF (grant AST18-1546).


\section*{Data Availability}
Data available upon reasonable request.



\bibliographystyle{mnras}
\bibliography{paper} 

\begin{thebibliography}{}
\makeatletter
\relax
\def\mn@urlcharsother{\let\do\@makeother \do\$\do\&\do\#\do\^\do\_\do\%\do\~}
\def\mn@doi{\begingroup\mn@urlcharsother \@ifnextchar [ {\mn@doi@}
  {\mn@doi@[]}}
\def\mn@doi@[#1]#2{\def\@tempa{#1}\ifx\@tempa\@empty \href
  {http://dx.doi.org/#2} {doi:#2}\else \href {http://dx.doi.org/#2} {#1}\fi
  \endgroup}
\def\mn@eprint#1#2{\mn@eprint@#1:#2::\@nil}
\def\mn@eprint@arXiv#1{\href {http://arxiv.org/abs/#1} {{\tt arXiv:#1}}}
\def\mn@eprint@dblp#1{\href {http://dblp.uni-trier.de/rec/bibtex/#1.xml}
  {dblp:#1}}
\def\mn@eprint@#1:#2:#3:#4\@nil{\def\@tempa {#1}\def\@tempb {#2}\def\@tempc
  {#3}\ifx \@tempc \@empty \let \@tempc \@tempb \let \@tempb \@tempa \fi \ifx
  \@tempb \@empty \def\@tempb {arXiv}\fi \@ifundefined
  {mn@eprint@\@tempb}{\@tempb:\@tempc}{\expandafter \expandafter \csname
  mn@eprint@\@tempb\endcsname \expandafter{\@tempc}}}

\bibitem[\protect\citeauthoryear{{Abbott} et~al.,}{{Abbott}
  et~al.}{2021}]{des2021}
{Abbott} T.~M.~C.,  et~al., 2021, \mn@doi [\apjs] {10.3847/1538-4365/ac00b3},
  \href {https://ui.adsabs.harvard.edu/abs/2021ApJS..255...20A} {255, 20}

\bibitem[\protect\citeauthoryear{{Agertz}, {Kravtsov}, {Leitner}  \&
  {Gnedin}}{{Agertz} et~al.}{2013}]{agertz2013}
{Agertz} O.,  {Kravtsov} A.~V.,  {Leitner} S.~N.,   {Gnedin} N.~Y.,  2013,
  \mn@doi [\apj] {10.1088/0004-637X/770/1/25}, \href
  {https://ui.adsabs.harvard.edu/abs/2013ApJ...770...25A} {770, 25}

\bibitem[\protect\citeauthoryear{{Agertz} et~al.,}{{Agertz}
  et~al.}{2020}]{agertz2020}
{Agertz} O.,  et~al., 2020, \mn@doi [\mnras] {10.1093/mnras/stz3053}, \href
  {https://ui.adsabs.harvard.edu/abs/2020MNRAS.491.1656A} {491, 1656}

\bibitem[\protect\citeauthoryear{{Battaglia} \& {Nipoti}}{{Battaglia} \&
  {Nipoti}}{2022}]{battaglia2022}
{Battaglia} G.,  {Nipoti} C.,  2022, \mn@doi [Nature Astronomy]
  {10.1038/s41550-022-01638-7}, \href
  {https://ui.adsabs.harvard.edu/abs/2022NatAs...6..659B} {6, 659}

\bibitem[\protect\citeauthoryear{{Battaglia}, {Helmi}  \&
  {Breddels}}{{Battaglia} et~al.}{2013}]{battaglia2013}
{Battaglia} G.,  {Helmi} A.,   {Breddels} M.,  2013, \mn@doi [\nar]
  {10.1016/j.newar.2013.05.003}, \href
  {https://ui.adsabs.harvard.edu/abs/2013NewAR..57...52B} {57, 52}

\bibitem[\protect\citeauthoryear{{Bechtol} et~al.,}{{Bechtol}
  et~al.}{2015}]{betchol2015}
{Bechtol} K.,  et~al., 2015, \mn@doi [\apj] {10.1088/0004-637X/807/1/50}, \href
  {https://ui.adsabs.harvard.edu/abs/2015ApJ...807...50B} {807, 50}

\bibitem[\protect\citeauthoryear{{Belokurov}}{{Belokurov}}{2013}]{belokurov2013}
{Belokurov} V.,  2013, \mn@doi [\nar] {10.1016/j.newar.2013.07.001}, \href
  {https://ui.adsabs.harvard.edu/abs/2013NewAR..57..100B} {57, 100}

\bibitem[\protect\citeauthoryear{{Belokurov} et~al.,}{{Belokurov}
  et~al.}{2007}]{belokurov2007}
{Belokurov} V.,  et~al., 2007, \mn@doi [\apj] {10.1086/509718}, \href
  {https://ui.adsabs.harvard.edu/abs/2007ApJ...654..897B} {654, 897}

\bibitem[\protect\citeauthoryear{{Belokurov} et~al.,}{{Belokurov}
  et~al.}{2008}]{belokurov2008}
{Belokurov} V.,  et~al., 2008, \mn@doi [\apjl] {10.1086/592962}, \href
  {https://ui.adsabs.harvard.edu/abs/2008ApJ...686L..83B} {686, L83}

\bibitem[\protect\citeauthoryear{{Belokurov} et~al.,}{{Belokurov}
  et~al.}{2009}]{belokurov2009}
{Belokurov} V.,  et~al., 2009, \mn@doi [\mnras]
  {10.1111/j.1365-2966.2009.15106.x}, \href
  {https://ui.adsabs.harvard.edu/abs/2009MNRAS.397.1748B} {397, 1748}

\bibitem[\protect\citeauthoryear{{Belokurov} et~al.,}{{Belokurov}
  et~al.}{2010}]{belokurov2010}
{Belokurov} V.,  et~al., 2010, \mn@doi [\apjl] {10.1088/2041-8205/712/1/L103},
  \href {https://ui.adsabs.harvard.edu/abs/2010ApJ...712L.103B} {712, L103}

\bibitem[\protect\citeauthoryear{Benitez-Llambay}{Benitez-Llambay}{2015}]{alejandro_benitez_llambay_2015_21703}
Benitez-Llambay A.,  2015, py-sphviewer: Py-SPHViewer v1.0.0,
  \mn@doi{10.5281/zenodo.21703}, \url {http://dx.doi.org/10.5281/zenodo.21703}

\bibitem[\protect\citeauthoryear{{Binney} \& {Tremaine}}{{Binney} \&
  {Tremaine}}{2008}]{binney2008}
{Binney} J.,  {Tremaine} S.,  2008, {Galactic Dynamics: Second Edition}

\bibitem[\protect\citeauthoryear{{Bromm} \& {Yoshida}}{{Bromm} \&
  {Yoshida}}{2011}]{bromm2011}
{Bromm} V.,  {Yoshida} N.,  2011, \mn@doi [\araa]
  {10.1146/annurev-astro-081710-102608}, \href
  {https://ui.adsabs.harvard.edu/abs/2011ARA&A..49..373B} {49, 373}

\bibitem[\protect\citeauthoryear{{Brown} et~al.,}{{Brown}
  et~al.}{2014}]{brown2014}
{Brown} T.~M.,  et~al., 2014, \mn@doi [\apj] {10.1088/0004-637X/796/2/91},
  \href {https://ui.adsabs.harvard.edu/abs/2014ApJ...796...91B} {796, 91}

\bibitem[\protect\citeauthoryear{{Bullock} \& {Boylan-Kolchin}}{{Bullock} \&
  {Boylan-Kolchin}}{2017}]{bb2017}
{Bullock} J.~S.,  {Boylan-Kolchin} M.,  2017, \mn@doi [\araa]
  {10.1146/annurev-astro-091916-055313}, \href
  {https://ui.adsabs.harvard.edu/abs/2017ARA&A..55..343B} {55, 343}

\bibitem[\protect\citeauthoryear{{Cerny} et~al.,}{{Cerny}
  et~al.}{2021}]{cerny2021}
{Cerny} W.,  et~al., 2021, \mn@doi [\apjl] {10.3847/2041-8213/ac2d9a}, \href
  {https://ui.adsabs.harvard.edu/abs/2021ApJ...920L..44C} {920, L44}

\bibitem[\protect\citeauthoryear{{Chabrier}}{{Chabrier}}{2003}]{chabrier2003}
{Chabrier} G.,  2003, \mn@doi [\pasp] {10.1086/376392}, \href
  {https://ui.adsabs.harvard.edu/abs/2003PASP..115..763C} {115, 763}

\bibitem[\protect\citeauthoryear{{Chiti} et~al.,}{{Chiti}
  et~al.}{2021}]{chiti2021}
{Chiti} A.,  et~al., 2021, \mn@doi [Nature Astronomy]
  {10.1038/s41550-020-01285-w}, \href
  {https://ui.adsabs.harvard.edu/abs/2021NatAs...5..392C} {5, 392}

\bibitem[\protect\citeauthoryear{{Chiti} et~al.,}{{Chiti}
  et~al.}{2022}]{chiti2022}
{Chiti} A.,  et~al., 2022, arXiv e-prints, \href
  {https://ui.adsabs.harvard.edu/abs/2022arXiv220501740C} {p. arXiv:2205.01740}

\bibitem[\protect\citeauthoryear{{Collins} \& {Read}}{{Collins} \&
  {Read}}{2022}]{collins2022}
{Collins} M. L.~M.,  {Read} J.~I.,  2022, \mn@doi [Nature Astronomy]
  {10.1038/s41550-022-01657-4}, \href
  {https://ui.adsabs.harvard.edu/abs/2022NatAs.tmp..105C} {}

\bibitem[\protect\citeauthoryear{{Collins}, {Tollerud}, {Sand}, {Bonaca},
  {Willman}  \& {Strader}}{{Collins} et~al.}{2017}]{collins2017}
{Collins} M. L.~M.,  {Tollerud} E.~J.,  {Sand} D.~J.,  {Bonaca} A.,  {Willman}
  B.,   {Strader} J.,  2017, \mn@doi [\mnras] {10.1093/mnras/stx067}, \href
  {https://ui.adsabs.harvard.edu/abs/2017MNRAS.467..573C} {467, 573}

\bibitem[\protect\citeauthoryear{{Collins}, {Charles}, {Mart{\'\i}nez-Delgado},
  {Monelli}, {Karim}, {Donatiello}, {Tollerud}  \& {Boschin}}{{Collins}
  et~al.}{2022}]{collinspegV}
{Collins} M. L.~M.,  {Charles} E. J.~E.,  {Mart{\'\i}nez-Delgado} D.,
  {Monelli} M.,  {Karim} N.,  {Donatiello} G.,  {Tollerud} E.~J.,   {Boschin}
  W.,  2022, arXiv e-prints, \href
  {https://ui.adsabs.harvard.edu/abs/2022arXiv220409068C} {p. arXiv:2204.09068}

\bibitem[\protect\citeauthoryear{{Crnojevi{\'c}}, {Sand}, {Zaritsky},
  {Spekkens}, {Willman}  \& {Hargis}}{{Crnojevi{\'c}}
  et~al.}{2016}]{eritwo2016}
{Crnojevi{\'c}} D.,  {Sand} D.~J.,  {Zaritsky} D.,  {Spekkens} K.,  {Willman}
  B.,   {Hargis} J.~R.,  2016, \mn@doi [\apjl] {10.3847/2041-8205/824/1/L14},
  \href {https://ui.adsabs.harvard.edu/abs/2016ApJ...824L..14C} {824, L14}

\bibitem[\protect\citeauthoryear{{Deason}, {Bose}, {Fattahi}, {Amorisco},
  {Hellwing}  \& {Frenk}}{{Deason} et~al.}{2022}]{deason2022}
{Deason} A.~J.,  {Bose} S.,  {Fattahi} A.,  {Amorisco} N.~C.,  {Hellwing} W.,
  {Frenk} C.~S.,  2022, \mn@doi [\mnras] {10.1093/mnras/stab3524}, \href
  {https://ui.adsabs.harvard.edu/abs/2022MNRAS.511.4044D} {511, 4044}

\bibitem[\protect\citeauthoryear{{Drlica-Wagner} et~al.,}{{Drlica-Wagner}
  et~al.}{2015}]{dw2015}
{Drlica-Wagner} A.,  et~al., 2015, \mn@doi [\apj]
  {10.1088/0004-637X/813/2/109}, \href
  {https://ui.adsabs.harvard.edu/abs/2015ApJ...813..109D} {813, 109}

\bibitem[\protect\citeauthoryear{{Drlica-Wagner} et~al.,}{{Drlica-Wagner}
  et~al.}{2016}]{dw2016}
{Drlica-Wagner} A.,  et~al., 2016, \mn@doi [\apjl]
  {10.3847/2041-8205/833/1/L5}, \href
  {https://ui.adsabs.harvard.edu/abs/2016ApJ...833L...5D} {833, L5}

\bibitem[\protect\citeauthoryear{{Drlica-Wagner} et~al.,}{{Drlica-Wagner}
  et~al.}{2021}]{dw2021}
{Drlica-Wagner} A.,  et~al., 2021, \mn@doi [\apjs] {10.3847/1538-4365/ac079d},
  \href {https://ui.adsabs.harvard.edu/abs/2021ApJS..256....2D} {256, 2}

\bibitem[\protect\citeauthoryear{{Eisenstein} \& {Hut}}{{Eisenstein} \&
  {Hut}}{1998}]{1998ApJ...498..137E}
{Eisenstein} D.~J.,  {Hut} P.,  1998, \mn@doi [\apj] {10.1086/305535}, \href
  {https://ui.adsabs.harvard.edu/abs/1998ApJ...498..137E} {498, 137}

\bibitem[\protect\citeauthoryear{{Filion} \& {Wyse}}{{Filion} \&
  {Wyse}}{2021}]{filionwyse2021}
{Filion} C.,  {Wyse} R. F.~G.,  2021, \mn@doi [\apj]
  {10.3847/1538-4357/ac2df1}, \href
  {https://ui.adsabs.harvard.edu/abs/2021ApJ...923..218F} {923, 218}

\bibitem[\protect\citeauthoryear{{Flaugher} et~al.,}{{Flaugher}
  et~al.}{2015}]{decam2015}
{Flaugher} B.,  et~al., 2015, \mn@doi [\aj] {10.1088/0004-6256/150/5/150},
  \href {https://ui.adsabs.harvard.edu/abs/2015AJ....150..150F} {150, 150}

\bibitem[\protect\citeauthoryear{{Foreman-Mackey}, {Hogg}, {Lang}  \&
  {Goodman}}{{Foreman-Mackey} et~al.}{2013}]{foreman2013}
{Foreman-Mackey} D.,  {Hogg} D.~W.,  {Lang} D.,   {Goodman} J.,  2013, \mn@doi
  [\pasp] {10.1086/670067}, \href
  {https://ui.adsabs.harvard.edu/abs/2013PASP..125..306F} {125, 306}

\bibitem[\protect\citeauthoryear{{Frebel}, {Simon}  \& {Kirby}}{{Frebel}
  et~al.}{2014}]{frebel2014}
{Frebel} A.,  {Simon} J.~D.,   {Kirby} E.~N.,  2014, \mn@doi [\apj]
  {10.1088/0004-637X/786/1/74}, \href
  {https://ui.adsabs.harvard.edu/abs/2014ApJ...786...74F} {786, 74}

\bibitem[\protect\citeauthoryear{{Fritz}, {Battaglia}, {Pawlowski},
  {Kallivayalil}, {van der Marel}, {Sohn}, {Brook}  \& {Besla}}{{Fritz}
  et~al.}{2018}]{fritz2018}
{Fritz} T.~K.,  {Battaglia} G.,  {Pawlowski} M.~S.,  {Kallivayalil} N.,  {van
  der Marel} R.,  {Sohn} S.~T.,  {Brook} C.,   {Besla} G.,  2018, \mn@doi
  [\aap] {10.1051/0004-6361/201833343}, \href
  {https://ui.adsabs.harvard.edu/abs/2018A&A...619A.103F} {619, A103}

\bibitem[\protect\citeauthoryear{{Gaia Collaboration} et~al.,}{{Gaia
  Collaboration} et~al.}{2021}]{gaiaedr32021}
{Gaia Collaboration} et~al., 2021, \mn@doi [\aap]
  {10.1051/0004-6361/202039657}, \href
  {https://ui.adsabs.harvard.edu/abs/2021A&A...649A...1G} {649, A1}

\bibitem[\protect\citeauthoryear{{Genina}, {Read}, {Fattahi}  \&
  {Frenk}}{{Genina} et~al.}{2022}]{genina2022}
{Genina} A.,  {Read} J.~I.,  {Fattahi} A.,   {Frenk} C.~S.,  2022, \mn@doi
  [\mnras] {10.1093/mnras/stab3526}, \href
  {https://ui.adsabs.harvard.edu/abs/2022MNRAS.510.2186G} {510, 2186}

\bibitem[\protect\citeauthoryear{{Haardt} \& {Madau}}{{Haardt} \&
  {Madau}}{1996}]{haardt1996}
{Haardt} F.,  {Madau} P.,  1996, \mn@doi [\apj] {10.1086/177035}, \href
  {https://ui.adsabs.harvard.edu/abs/1996ApJ...461...20H} {461, 20}

\bibitem[\protect\citeauthoryear{{Homma} et~al.,}{{Homma}
  et~al.}{2016}]{homma2016}
{Homma} D.,  et~al., 2016, \mn@doi [\apj] {10.3847/0004-637X/832/1/21}, \href
  {https://ui.adsabs.harvard.edu/abs/2016ApJ...832...21H} {832, 21}

\bibitem[\protect\citeauthoryear{{Homma} et~al.,}{{Homma}
  et~al.}{2018}]{Homma2018}
{Homma} D.,  et~al., 2018, \mn@doi [\pasj] {10.1093/pasj/psx050}, \href
  {https://ui.adsabs.harvard.edu/abs/2018PASJ...70S..18H} {70, S18}

\bibitem[\protect\citeauthoryear{{Ibata} et~al.,}{{Ibata}
  et~al.}{2014}]{pandas2014}
{Ibata} R.~A.,  et~al., 2014, \mn@doi [\apj] {10.1088/0004-637X/780/2/128},
  \href {https://ui.adsabs.harvard.edu/abs/2014ApJ...780..128I} {780, 128}

\bibitem[\protect\citeauthoryear{{Irwin} \& {Hatzidimitriou}}{{Irwin} \&
  {Hatzidimitriou}}{1995}]{irwin1995}
{Irwin} M.,  {Hatzidimitriou} D.,  1995, \mn@doi [\mnras]
  {10.1093/mnras/277.4.1354}, \href
  {https://ui.adsabs.harvard.edu/abs/1995MNRAS.277.1354I} {277, 1354}

\bibitem[\protect\citeauthoryear{{Kahn}}{{Kahn}}{2018}]{lsst2018}
{Kahn} S.,  2018, in 42nd COSPAR Scientific Assembly. pp E1.16--5--18

\bibitem[\protect\citeauthoryear{{Katz} \& {White}}{{Katz} \&
  {White}}{1993}]{katz1993}
{Katz} N.,  {White} S. D.~M.,  1993, \mn@doi [\apj] {10.1086/172935}, \href
  {https://ui.adsabs.harvard.edu/abs/1993ApJ...412..455K} {412, 455}

\bibitem[\protect\citeauthoryear{{Kim} \& {Jerjen}}{{Kim} \&
  {Jerjen}}{2015}]{kj2015}
{Kim} D.,  {Jerjen} H.,  2015, \mn@doi [\apjl] {10.1088/2041-8205/808/2/L39},
  \href {https://ui.adsabs.harvard.edu/abs/2015ApJ...808L..39K} {808, L39}

\bibitem[\protect\citeauthoryear{{Kim}, {Jerjen}, {Milone}, {Mackey}  \& {Da
  Costa}}{{Kim} et~al.}{2015a}]{kim2015b}
{Kim} D.,  {Jerjen} H.,  {Milone} A.~P.,  {Mackey} D.,   {Da Costa} G.~S.,
  2015a, \mn@doi [\apj] {10.1088/0004-637X/803/2/63}, \href
  {https://ui.adsabs.harvard.edu/abs/2015ApJ...803...63K} {803, 63}

\bibitem[\protect\citeauthoryear{{Kim}, {Jerjen}, {Mackey}, {Da Costa}  \&
  {Milone}}{{Kim} et~al.}{2015b}]{kim2015a}
{Kim} D.,  {Jerjen} H.,  {Mackey} D.,  {Da Costa} G.~S.,   {Milone} A.~P.,
  2015b, \mn@doi [\apjl] {10.1088/2041-8205/804/2/L44}, \href
  {https://ui.adsabs.harvard.edu/abs/2015ApJ...804L..44K} {804, L44}

\bibitem[\protect\citeauthoryear{{Kimm}, {Cen}, {Devriendt}, {Dubois}  \&
  {Slyz}}{{Kimm} et~al.}{2015}]{kimm2014}
{Kimm} T.,  {Cen} R.,  {Devriendt} J.,  {Dubois} Y.,   {Slyz} A.,  2015,
  \mn@doi [\mnras] {10.1093/mnras/stv1211}, \href
  {https://ui.adsabs.harvard.edu/abs/2015MNRAS.451.2900K} {451, 2900}

\bibitem[\protect\citeauthoryear{{Kirby}, {Cohen}, {Guhathakurta}, {Cheng},
  {Bullock}  \& {Gallazzi}}{{Kirby} et~al.}{2013}]{kirby2013}
{Kirby} E.~N.,  {Cohen} J.~G.,  {Guhathakurta} P.,  {Cheng} L.,  {Bullock}
  J.~S.,   {Gallazzi} A.,  2013, \mn@doi [\apj] {10.1088/0004-637X/779/2/102},
  \href {https://ui.adsabs.harvard.edu/abs/2013ApJ...779..102K} {779, 102}

\bibitem[\protect\citeauthoryear{{Koposov} et~al.,}{{Koposov}
  et~al.}{2008}]{koposov2008}
{Koposov} S.,  et~al., 2008, \mn@doi [\apj] {10.1086/589911}, \href
  {https://ui.adsabs.harvard.edu/abs/2008ApJ...686..279K} {686, 279}

\bibitem[\protect\citeauthoryear{{Koposov}, {Belokurov}, {Torrealba}  \&
  {Evans}}{{Koposov} et~al.}{2015}]{koposov2015}
{Koposov} S.~E.,  {Belokurov} V.,  {Torrealba} G.,   {Evans} N.~W.,  2015,
  \mn@doi [\apj] {10.1088/0004-637X/805/2/130}, \href
  {https://ui.adsabs.harvard.edu/abs/2015ApJ...805..130K} {805, 130}

\bibitem[\protect\citeauthoryear{{Koposov} et~al.,}{{Koposov}
  et~al.}{2018}]{koposov2018}
{Koposov} S.~E.,  et~al., 2018, \mn@doi [\mnras] {10.1093/mnras/sty1772}, \href
  {https://ui.adsabs.harvard.edu/abs/2018MNRAS.479.5343K} {479, 5343}

\bibitem[\protect\citeauthoryear{{K{\"u}pper}, {Johnston}, {Mieske}, {Collins}
  \& {Tollerud}}{{K{\"u}pper} et~al.}{2017}]{kupper2017}
{K{\"u}pper} A. H.~W.,  {Johnston} K.~V.,  {Mieske} S.,  {Collins} M. L.~M.,
  {Tollerud} E.~J.,  2017, \mn@doi [\apj] {10.3847/1538-4357/834/2/112}, \href
  {https://ui.adsabs.harvard.edu/abs/2017ApJ...834..112K} {834, 112}

\bibitem[\protect\citeauthoryear{{Laevens} et~al.,}{{Laevens}
  et~al.}{2015}]{Laevens2015}
{Laevens} B. P.~M.,  et~al., 2015, \mn@doi [\apj] {10.1088/0004-637X/813/1/44},
  \href {https://ui.adsabs.harvard.edu/abs/2015ApJ...813...44L} {813, 44}

\bibitem[\protect\citeauthoryear{{Li} et~al.,}{{Li} et~al.}{2018}]{li2018}
{Li} T.~S.,  et~al., 2018, \mn@doi [\apj] {10.3847/1538-4357/aadf91}, \href
  {https://ui.adsabs.harvard.edu/abs/2018ApJ...866...22L} {866, 22}

\bibitem[\protect\citeauthoryear{{{\L}okas}, {Gajda}  \&
  {Kazantzidis}}{{{\L}okas} et~al.}{2013}]{lokas2013}
{{\L}okas} E.~L.,  {Gajda} G.,   {Kazantzidis} S.,  2013, \mn@doi [\mnras]
  {10.1093/mnras/stt774}, \href
  {https://ui.adsabs.harvard.edu/abs/2013MNRAS.433..878L} {433, 878}

\bibitem[\protect\citeauthoryear{{Longeard} et~al.,}{{Longeard}
  et~al.}{2021}]{pristine2021}
{Longeard} N.,  et~al., 2021, arXiv e-prints, \href
  {https://ui.adsabs.harvard.edu/abs/2021arXiv210710849L} {p. arXiv:2107.10849}

\bibitem[\protect\citeauthoryear{{Longeard} et~al.,}{{Longeard}
  et~al.}{2022}]{longeard2022}
{Longeard} N.,  et~al., 2022, \mn@doi [\mnras] {10.1093/mnras/stac1827}, \href
  {https://ui.adsabs.harvard.edu/abs/2022MNRAS.516.2348L} {516, 2348}

\bibitem[\protect\citeauthoryear{{Longeard} et~al.,}{{Longeard}
  et~al.}{2023}]{longeard2023}
{Longeard} N.,  et~al., 2023, \mn@doi [arXiv e-prints]
  {10.48550/arXiv.2304.13046}, \href
  {https://ui.adsabs.harvard.edu/abs/2023arXiv230413046L} {p. arXiv:2304.13046}

\bibitem[\protect\citeauthoryear{{Martin}, {de Jong}  \& {Rix}}{{Martin}
  et~al.}{2008}]{martin2008}
{Martin} N.~F.,  {de Jong} J. T.~A.,   {Rix} H.-W.,  2008, \mn@doi [\apj]
  {10.1086/590336}, \href
  {https://ui.adsabs.harvard.edu/abs/2008ApJ...684.1075M} {684, 1075}

\bibitem[\protect\citeauthoryear{{Martin} et~al.,}{{Martin}
  et~al.}{2015}]{martin2015}
{Martin} N.~F.,  et~al., 2015, \mn@doi [\apjl] {10.1088/2041-8205/804/1/L5},
  \href {https://ui.adsabs.harvard.edu/abs/2015ApJ...804L...5M} {804, L5}

\bibitem[\protect\citeauthoryear{{Martin} et~al.,}{{Martin}
  et~al.}{2016}]{martin2016}
{Martin} N.~F.,  et~al., 2016, \mn@doi [\apj] {10.3847/1538-4357/833/2/167},
  \href {https://ui.adsabs.harvard.edu/abs/2016ApJ...833..167M} {833, 167}

\bibitem[\protect\citeauthoryear{{Mazzarini}, {Just}, {Macci{\`o}}  \&
  {Moetazedian}}{{Mazzarini} et~al.}{2020}]{mazzarini2020}
{Mazzarini} M.,  {Just} A.,  {Macci{\`o}} A.~V.,   {Moetazedian} R.,  2020,
  \mn@doi [\aap] {10.1051/0004-6361/202037558}, \href
  {https://ui.adsabs.harvard.edu/abs/2020A&A...636A.106M} {636, A106}

\bibitem[\protect\citeauthoryear{{McConnachie}}{{McConnachie}}{2012}]{mcconachie2012}
{McConnachie} A.~W.,  2012, \mn@doi [\aj] {10.1088/0004-6256/144/1/4}, \href
  {https://ui.adsabs.harvard.edu/abs/2012AJ....144....4M} {144, 4}

\bibitem[\protect\citeauthoryear{{McConnachie} \& {Venn}}{{McConnachie} \&
  {Venn}}{2020}]{mcconachie2020}
{McConnachie} A.~W.,  {Venn} K.~A.,  2020, \mn@doi [Research Notes of the
  American Astronomical Society] {10.3847/2515-5172/abd18b}, \href
  {https://ui.adsabs.harvard.edu/abs/2020RNAAS...4..229M} {4, 229}

\bibitem[\protect\citeauthoryear{{McConnachie}, {Arimoto}, {Irwin}  \&
  {Tolstoy}}{{McConnachie} et~al.}{2006}]{mcconachie2006}
{McConnachie} A.~W.,  {Arimoto} N.,  {Irwin} M.,   {Tolstoy} E.,  2006, \mn@doi
  [\mnras] {10.1111/j.1365-2966.2006.11053.x}, \href
  {https://ui.adsabs.harvard.edu/abs/2006MNRAS.373..715M} {373, 715}

\bibitem[\protect\citeauthoryear{{McQuinn} et~al.,}{{McQuinn}
  et~al.}{2015}]{leop2015}
{McQuinn} K. B.~W.,  et~al., 2015, \mn@doi [\apj]
  {10.1088/0004-637X/812/2/158}, \href
  {https://ui.adsabs.harvard.edu/abs/2015ApJ...812..158M} {812, 158}

\bibitem[\protect\citeauthoryear{{Mu{\~n}oz}, {Geha}  \& {Willman}}{{Mu{\~n}oz}
  et~al.}{2010}]{munoz2010}
{Mu{\~n}oz} R.~R.,  {Geha} M.,   {Willman} B.,  2010, \mn@doi [\aj]
  {10.1088/0004-6256/140/1/138}, \href
  {https://ui.adsabs.harvard.edu/abs/2010AJ....140..138M} {140, 138}

\bibitem[\protect\citeauthoryear{{Mu{\~n}oz}, {Padmanabhan}  \&
  {Geha}}{{Mu{\~n}oz} et~al.}{2012}]{munoz2012}
{Mu{\~n}oz} R.~R.,  {Padmanabhan} N.,   {Geha} M.,  2012, \mn@doi [\apj]
  {10.1088/0004-637X/745/2/127}, \href
  {https://ui.adsabs.harvard.edu/abs/2012ApJ...745..127M} {745, 127}

\bibitem[\protect\citeauthoryear{{Mutlu-Pakdil} et~al.,}{{Mutlu-Pakdil}
  et~al.}{2018}]{rettwo}
{Mutlu-Pakdil} B.,  et~al., 2018, \mn@doi [\apj] {10.3847/1538-4357/aacd0e},
  \href {https://ui.adsabs.harvard.edu/abs/2018ApJ...863...25M} {863, 25}

\bibitem[\protect\citeauthoryear{{Mutlu-Pakdil} et~al.,}{{Mutlu-Pakdil}
  et~al.}{2019}]{mutlu-pakdil2019}
{Mutlu-Pakdil} B.,  et~al., 2019, \mn@doi [\apj] {10.3847/1538-4357/ab45ec},
  \href {https://ui.adsabs.harvard.edu/abs/2019ApJ...885...53M} {885, 53}

\bibitem[\protect\citeauthoryear{{O{\~n}orbe}, {Garrison-Kimmel}, {Maller},
  {Bullock}, {Rocha}  \& {Hahn}}{{O{\~n}orbe} et~al.}{2014}]{onorbe2014zoom}
{O{\~n}orbe} J.,  {Garrison-Kimmel} S.,  {Maller} A.~H.,  {Bullock} J.~S.,
  {Rocha} M.,   {Hahn} O.,  2014, \mn@doi [\mnras] {10.1093/mnras/stt2020},
  \href {https://ui.adsabs.harvard.edu/abs/2014MNRAS.437.1894O} {437, 1894}

\bibitem[\protect\citeauthoryear{{Pace}, {Erkal}  \& {Li}}{{Pace}
  et~al.}{2022}]{pace2022}
{Pace} A.~B.,  {Erkal} D.,   {Li} T.~S.,  2022, arXiv e-prints, \href
  {https://ui.adsabs.harvard.edu/abs/2022arXiv220505699P} {p. arXiv:2205.05699}

\bibitem[\protect\citeauthoryear{{Planck Collaboration} et~al.,}{{Planck
  Collaboration} et~al.}{2014}]{planck2014}
{Planck Collaboration} et~al., 2014, \mn@doi [\aap]
  {10.1051/0004-6361/201321591}, \href
  {https://ui.adsabs.harvard.edu/abs/2014A&A...571A..16P} {571, A16}

\bibitem[\protect\citeauthoryear{{Pontzen} \& {Tremmel}}{{Pontzen} \&
  {Tremmel}}{2018}]{pontzen2018}
{Pontzen} A.,  {Tremmel} M.,  2018, \mn@doi [\apjs] {10.3847/1538-4365/aac832},
  \href {https://ui.adsabs.harvard.edu/abs/2018ApJS..237...23P} {237, 23}

\bibitem[\protect\citeauthoryear{{Pontzen}, {Ro{\v{s}}kar}, {Stinson}  \&
  {Woods}}{{Pontzen} et~al.}{2013}]{pontzen2013}
{Pontzen} A.,  {Ro{\v{s}}kar} R.,  {Stinson} G.,   {Woods} R.,  2013, {pynbody:
  N-Body/SPH analysis for python}, Astrophysics Source Code Library, record
  ascl:1305.002 (\mn@eprint {ascl} {1305.002})

\bibitem[\protect\citeauthoryear{{Pozo}, {Broadhurst}, {Emami}  \&
  {Smoot}}{{Pozo} et~al.}{2022}]{pozo2022}
{Pozo} A.,  {Broadhurst} T.,  {Emami} R.,   {Smoot} G.,  2022, \mn@doi [\mnras]
  {10.1093/mnras/stac1862}, \href
  {https://ui.adsabs.harvard.edu/abs/2022MNRAS.515.2624P} {515, 2624}

\bibitem[\protect\citeauthoryear{{Read}, {Wilkinson}, {Evans}, {Gilmore}  \&
  {Kleyna}}{{Read} et~al.}{2006}]{read2006c}
{Read} J.~I.,  {Wilkinson} M.~I.,  {Evans} N.~W.,  {Gilmore} G.,   {Kleyna}
  J.~T.,  2006, \mn@doi [\mnras] {10.1111/j.1365-2966.2005.09861.x}, \href
  {https://ui.adsabs.harvard.edu/abs/2006MNRAS.366..429R} {366, 429}

\bibitem[\protect\citeauthoryear{{Rey}, {Pontzen}, {Agertz}, {Orkney}, {Read},
  {Saintonge}  \& {Pedersen}}{{Rey} et~al.}{2019}]{rey2019}
{Rey} M.~P.,  {Pontzen} A.,  {Agertz} O.,  {Orkney} M. D.~A.,  {Read} J.~I.,
  {Saintonge} A.,   {Pedersen} C.,  2019, \mn@doi [\apjl]
  {10.3847/2041-8213/ab53dd}, \href
  {https://ui.adsabs.harvard.edu/abs/2019ApJ...886L...3R} {886, L3}

\bibitem[\protect\citeauthoryear{{Rey}, {Pontzen}, {Agertz}, {Orkney}, {Read}
  \& {Rosdahl}}{{Rey} et~al.}{2020}]{rey2020}
{Rey} M.~P.,  {Pontzen} A.,  {Agertz} O.,  {Orkney} M. D.~A.,  {Read} J.~I.,
  {Rosdahl} J.,  2020, \mn@doi [\mnras] {10.1093/mnras/staa1640}, \href
  {https://ui.adsabs.harvard.edu/abs/2020MNRAS.497.1508R} {497, 1508}

\bibitem[\protect\citeauthoryear{{Rey}, {Pontzen}, {Agertz}, {Orkney}, {Read},
  {Saintonge}, {Kim}  \& {Das}}{{Rey} et~al.}{2022}]{rey_edge_2021}
{Rey} M.~P.,  {Pontzen} A.,  {Agertz} O.,  {Orkney} M. D.~A.,  {Read} J.~I.,
  {Saintonge} A.,  {Kim} S.~Y.,   {Das} P.,  2022, \mn@doi [\mnras]
  {10.1093/mnras/stac502}, \href
  {https://ui.adsabs.harvard.edu/abs/2022MNRAS.511.5672R} {511, 5672}

\bibitem[\protect\citeauthoryear{{Sales}, {Wetzel}  \& {Fattahi}}{{Sales}
  et~al.}{2022}]{sales2022}
{Sales} L.~V.,  {Wetzel} A.,   {Fattahi} A.,  2022, \mn@doi [Nature Astronomy]
  {10.1038/s41550-022-01689-w}, \href
  {https://ui.adsabs.harvard.edu/abs/2022NatAs.tmp..130S} {}

\bibitem[\protect\citeauthoryear{{Sand}}{{Sand}}{2017}]{sand2017}
{Sand} D.,  2017, {The Origin of Ultra-Faint Galaxies}, HST Proposal id.15182.
  Cycle 25

\bibitem[\protect\citeauthoryear{{Sand}, {Olszewski}, {Willman}, {Zaritsky},
  {Seth}, {Harris}, {Piatek}  \& {Saha}}{{Sand} et~al.}{2009}]{sand2009}
{Sand} D.~J.,  {Olszewski} E.~W.,  {Willman} B.,  {Zaritsky} D.,  {Seth} A.,
  {Harris} J.,  {Piatek} S.,   {Saha} A.,  2009, \mn@doi [\apj]
  {10.1088/0004-637X/704/2/898}, \href
  {https://ui.adsabs.harvard.edu/abs/2009ApJ...704..898S} {704, 898}

\bibitem[\protect\citeauthoryear{{Sand} et~al.,}{{Sand}
  et~al.}{2022}]{sand2022}
{Sand} D.~J.,  et~al., 2022, arXiv e-prints, \href
  {https://ui.adsabs.harvard.edu/abs/2022arXiv220509129S} {p. arXiv:2205.09129}

\bibitem[\protect\citeauthoryear{{Schmidt}}{{Schmidt}}{1959}]{schmidt1959}
{Schmidt} M.,  1959, \mn@doi [\apj] {10.1086/146614}, \href
  {https://ui.adsabs.harvard.edu/abs/1959ApJ...129..243S} {129, 243}

\bibitem[\protect\citeauthoryear{{Sestito} et~al.,}{{Sestito}
  et~al.}{2023a}]{sestito2023}
{Sestito} F.,  et~al., 2023a, \mn@doi [arXiv e-prints]
  {10.48550/arXiv.2301.13214}, \href
  {https://ui.adsabs.harvard.edu/abs/2023arXiv230113214S} {p. arXiv:2301.13214}

\bibitem[\protect\citeauthoryear{{Sestito}, {Roediger}, {Navarro}, {Jensen},
  {Venn}, {Smith}, {Hayes}  \& {McConnachie}}{{Sestito}
  et~al.}{2023b}]{sestito2023b}
{Sestito} F.,  {Roediger} J.,  {Navarro} J.~F.,  {Jensen} J.,  {Venn} K.~A.,
  {Smith} S. E.~T.,  {Hayes} C.,   {McConnachie} A.~W.,  2023b, \mn@doi
  [\mnras] {10.1093/mnras/stad1417}, \href
  {https://ui.adsabs.harvard.edu/abs/2023MNRAS.523..123S} {523, 123}

\bibitem[\protect\citeauthoryear{{Shipp} et~al.,}{{Shipp}
  et~al.}{2023}]{shipp2023}
{Shipp} N.,  et~al., 2023, \mn@doi [\apj] {10.3847/1538-4357/acc582}, \href
  {https://ui.adsabs.harvard.edu/abs/2023ApJ...949...44S} {949, 44}

\bibitem[\protect\citeauthoryear{{Simon}}{{Simon}}{2018}]{simon2018}
{Simon} J.~D.,  2018, \mn@doi [\apj] {10.3847/1538-4357/aacdfb}, \href
  {https://ui.adsabs.harvard.edu/abs/2018ApJ...863...89S} {863, 89}

\bibitem[\protect\citeauthoryear{{Simon}}{{Simon}}{2019}]{simon2019}
{Simon} J.~D.,  2019, \mn@doi [\araa] {10.1146/annurev-astro-091918-104453},
  \href {https://ui.adsabs.harvard.edu/abs/2019ARA&A..57..375S} {57, 375}

\bibitem[\protect\citeauthoryear{{Stopyra}, {Pontzen}, {Peiris}, {Roth}  \&
  {Rey}}{{Stopyra} et~al.}{2021}]{stopyra2021}
{Stopyra} S.,  {Pontzen} A.,  {Peiris} H.,  {Roth} N.,   {Rey} M.~P.,  2021,
  \mn@doi [\apjs] {10.3847/1538-4365/abcd94}, \href
  {https://ui.adsabs.harvard.edu/abs/2021ApJS..252...28S} {252, 28}

\bibitem[\protect\citeauthoryear{{Tarumi}, {Yoshida}  \& {Frebel}}{{Tarumi}
  et~al.}{2021}]{tarumi2021}
{Tarumi} Y.,  {Yoshida} N.,   {Frebel} A.,  2021, \mn@doi [\apjl]
  {10.3847/2041-8213/ac024e}, \href
  {https://ui.adsabs.harvard.edu/abs/2021ApJ...914L..10T} {914, L10}

\bibitem[\protect\citeauthoryear{{Teyssier}}{{Teyssier}}{2002}]{teyssier2002}
{Teyssier} R.,  2002, \mn@doi [\aap] {10.1051/0004-6361:20011817}, \href
  {https://ui.adsabs.harvard.edu/abs/2002A&A...385..337T} {385, 337}

\bibitem[\protect\citeauthoryear{{Torrealba} et~al.,}{{Torrealba}
  et~al.}{2016}]{torrealba2016a}
{Torrealba} G.,  et~al., 2016, \mn@doi [\mnras] {10.1093/mnras/stw2051}, \href
  {https://ui.adsabs.harvard.edu/abs/2016MNRAS.463..712T} {463, 712}

\bibitem[\protect\citeauthoryear{{Torrealba} et~al.,}{{Torrealba}
  et~al.}{2018}]{torrealba2018}
{Torrealba} G.,  et~al., 2018, \mn@doi [\mnras] {10.1093/mnras/sty170}, \href
  {https://ui.adsabs.harvard.edu/abs/2018MNRAS.475.5085T} {475, 5085}

\bibitem[\protect\citeauthoryear{{Waller} et~al.,}{{Waller}
  et~al.}{2023}]{waller2023}
{Waller} F.,  et~al., 2023, \mn@doi [\mnras] {10.1093/mnras/stac3563}, \href
  {https://ui.adsabs.harvard.edu/abs/2023MNRAS.519.1349W} {519, 1349}

\bibitem[\protect\citeauthoryear{{Wheeler} et~al.,}{{Wheeler}
  et~al.}{2019}]{wheeler2019}
{Wheeler} C.,  et~al., 2019, \mn@doi [\mnras] {10.1093/mnras/stz2887}, \href
  {https://ui.adsabs.harvard.edu/abs/2019MNRAS.490.4447W} {490, 4447}

\bibitem[\protect\citeauthoryear{{Whiting}, {Hau}, {Irwin}  \&
  {Verdugo}}{{Whiting} et~al.}{2007}]{whiting2007}
{Whiting} A.~B.,  {Hau} G. K.~T.,  {Irwin} M.,   {Verdugo} M.,  2007, \mn@doi
  [\aj] {10.1086/510309}, \href
  {https://ui.adsabs.harvard.edu/abs/2007AJ....133..715W} {133, 715}

\bibitem[\protect\citeauthoryear{{Willman} et~al.,}{{Willman}
  et~al.}{2005}]{willman2005a}
{Willman} B.,  et~al., 2005, \mn@doi [\apjl] {10.1086/431760}, \href
  {https://ui.adsabs.harvard.edu/abs/2005ApJ...626L..85W} {626, L85}

\bibitem[\protect\citeauthoryear{{Yang}, {Hammer}, {Jiao}  \&
  {Pawlowski}}{{Yang} et~al.}{2022}]{yang2022}
{Yang} Y.,  {Hammer} F.,  {Jiao} Y.,   {Pawlowski} M.~S.,  2022, \mn@doi
  [\mnras] {10.1093/mnras/stac644}, \href
  {https://ui.adsabs.harvard.edu/abs/2022MNRAS.512.4171Y} {512, 4171}

\bibitem[\protect\citeauthoryear{{Zoutendijk} et~al.,}{{Zoutendijk}
  et~al.}{2021}]{zoutendijk2021}
{Zoutendijk} S.~L.,  et~al., 2021, arXiv e-prints, \href
  {https://ui.adsabs.harvard.edu/abs/2021arXiv211209374Z} {p. arXiv:2112.09374}

\bibitem[\protect\citeauthoryear{{de Jong}, {Martin}, {Rix}, {Smith}, {Jin}  \&
  {Macci{\`o}}}{{de Jong} et~al.}{2010}]{dejong2010}
{de Jong} J. T.~A.,  {Martin} N.~F.,  {Rix} H.-W.,  {Smith} K.~W.,  {Jin} S.,
  {Macci{\`o}} A.~V.,  2010, \mn@doi [\apj] {10.1088/0004-637X/710/2/1664},
  \href {https://ui.adsabs.harvard.edu/abs/2010ApJ...710.1664D} {710, 1664}

\bibitem[\protect\citeauthoryear{{von Hoerner}}{{von
  Hoerner}}{1957}]{hoerner1957}
{von Hoerner} S.,  1957, \mn@doi [\apj] {10.1086/146321}, \href
  {https://ui.adsabs.harvard.edu/abs/1957ApJ...125..451V} {125, 451}

\makeatother
\end{thebibliography}







\bsp	
\label{lastpage}
\end{document}